\documentclass[pre,twocolumn,superscriptaddress,floatfix,showpacs,showkeys,nofootinbib]{revtex4}
\usepackage{amsmath}
\usepackage{amssymb}
\usepackage{graphicx}
\usepackage{float}
\usepackage{color}
\newcommand\T{\rule{0pt}{2.6ex}}       
\newcommand{\bs}{\boldsymbol}
\newcommand{\ww}[1]{\underline{\underline{{\bf #1}}}}

\newcommand{\fcap}{F^{\text{Cap}}}

\newcommand{\be}{\begin{equation}}
\newcommand{\ee}{\end{equation}}
\newcommand{\bc}{\begin{center}}
\newcommand{\ec}{\end{center}}
\newcommand{\bea}{\begin{eqnarray}}
\newcommand{\eea}{\end{eqnarray}}
\newcommand{\ba}{\begin{aligned}}
\newcommand{\ea}{\end{aligned}}

\newcommand{\tr}{\text{tr}\hspace{.05em}}

\newcommand{\ave}[1]{\langle #1 \rangle}

\newcommand{\xoj}{x_{2}}

\newcommand{\ox}{_{1}}
\newcommand{\oy}{_{2}}

\newcommand{\oxx}{_{11}}
\newcommand{\oyy}{_{22}}
\newcommand{\ozz}{_{33}}
\newcommand{\oxy}{_{12}}

\newcommand{\scap}{\sigma^{\text{cap}}}
\newcommand{\sne}{\sigma^{\text{Ne}}}
\newcommand{\st}{\sigma^{\text{T}}}
\newcommand{\scon}{\sigma^{\text{c}}}
\newcommand{\sd}{\sigma^{\text{d}}}

\newcommand{\Fn}{F^{\text{N}}}

\newcommand{\Fnc}{F^{\text{N,c}}}

\newcommand{\Fc}{F^{\text{c}}}
\newcommand{\Fd}{F^{\text{d}}}
\newcommand{\zc}{z_{\text{c}}}
\newcommand{\zd}{z_{\text{d}}}

\newcommand{\cP}{\mathcal{P}}

\newcommand{\tauc}{\tau^{\text{c}}}
\newcommand{\tauz}{\tau_{\text{0}}}
\newcommand{\taui}{\tau^{\text{i}}}
\newcommand{\taucl}{\tau^{\text{cl}}}

\newcommand{\scl}{S^{\text{cl}}}

\begin{document}

\title{Flow of wet granular materials: a numerical study}

\author{Saeed Khamseh\footnote{Present address: 
Multiscale Mechanics (MSM), CTW, MESA+, University of Twente, P.O. Box 217, 7500
AE Enschede, The Netherlands}}
\email{s.khamseh-1@utwente.nl}
\affiliation{Universit\'e Paris-Est, Laboratoire Navier, 2 All\'ee Kepler, 77420 Champs-sur-Marne, France}

\author{Jean-No\"el Roux}
\email{jean-noel.roux@ifsttar.fr}
\affiliation{Universit\'e Paris-Est, Laboratoire Navier, 2 All\'ee Kepler, 77420 Champs-sur-Marne, France}

\author{Fran\c{c}ois Chevoir}
\email{francois.chevoir@ifsttar.fr}
\affiliation{Universit\'e Paris-Est, Laboratoire Navier, 2 All\'ee Kepler, 77420 Champs-sur-Marne, France}
\date{\today}

\begin{abstract}
We simulate dense assemblies of frictional spherical grains in steady shear flow under controlled normal stress P in the presence of a small 
amount of an interstitial liquid, which gives rise to capillary menisci, assumed isolated (pendular regime), and to attractive forces, which are 
hysteric: menisci form at contact, but do not break until
grains are separated by a finite rupture distance.
The system behavior depends on two dimensionless control parameters: inertial number $I$ and reduced pressure $P^* = aP/(\pi\Gamma)$, comparing confining forces
$\sim a^2P$ to meniscus tensile strength $F_0=\pi\Gamma a$, for grains of diameter $a$ joined by menisci with surface tension $\Gamma$. 
We pay special attention to the quasi-static limit of slow flow and observe systematic, enduring strain localization in some of the cohesion-dominated ($P^*\sim 0.1$) systems. 
Homogeneous steady flows are characterized by the dependence of internal friction coefficient $\mu^*$ and solid fraction $\Phi$ on $I$ and $P^*$. We also record 
normal stress differences, fairly small but not negligible, and increasing for decreasing $P^*$. 
The  system rheology is moderately sensitive to saturation within the pendular regime, but would be different in the absence of capillary hysteresis.
Capillary forces have a significant effect on the macroscopic behavior of the system, up to $P^*$ values of several units, especially for longer 
force ranges associated with larger menisci. The concept of effective 
pressure may be used to predict an order of magnitude for the strong increase of $\mu^*$  as $P^*$ decreases but such a crude approach is unable  to account for the complex structural changes induced by capillary cohesion, with a significant decrease of $\Phi$ and different agglomeration states and anisotropic fabric. Likewise, the Mohr-Coulomb criterion 
for pressure-dependent critical states is, at best, an approximation valid within a restricted range of pressures, with $P^*\ge 1$. At small enough $P^*$, large clusters of interacting grains form in slow flows, in which liquid bonds survive shear strains of several units. This affects the 
anisotropies associated to different interactions, and the shape of function $\mu^*(I)$, which departs more slowly from its quasistatic limit than
in cohesionless systems (possibly explaining the shear banding tendency).  
\end{abstract}

\maketitle

\section{Introduction}
Over the last decade, constitutive modelling  of dense granular flows has been proposed~\cite{Dacruz05,FoPo08} in terms of internal friction laws directly applying to normal stress-controlled steady shear flows, 
for which the internal state of the material is characterized by a single dimensionless number, the inertial parameter $I$~\cite{Gdr04}. 
Number $I$ might be regarded as a reduced, dimensionless form of shear rate $\dot\gamma=\partial v_1 / \partial x_2$, related to the stress $\sigma_{22}$ normal to flow
direction $x$ as $I = \dot{\gamma} \sqrt{\frac{m}{a\sigma_{22}}}$, $m$ denoting the particle mass and $a$ its diameter. The constitutive law relating the effective internal friction coefficient, $\mu^*$, 
defined as a stress ratio, $\mu^*= \sigma_{12}/\sigma_{22}$, to inertial number $I$ should be supplemented with a similar 
relation of solid fraction $\Phi$ to $I$~\cite{Dacruz05,Hatano07,PR08a,AzRa2014}. $I$ characterizes dynamical effects, and the quasistatic limit is that of vanishing $I$. 
In this limit of $I\to 0$, the material is in the so-called critical state of soil mechanics~\cite{DMWood} , i.e., quasistatic plastic shear flow at constant solid fraction $\Phi_c$, under constant stresses
 and effective internal friction $\mu^*_c$. In various experimental and numerical studies, the constitutive law, suitably generalized, was shown to apply to different grain shapes 
and flow geometries~\cite{JFP06,Koval09,ADRRC12}. A major advantage of the ``$\mu^*(I)$ and $\Phi(I)$'' approach is its ability to deal with both the quasistatic limit and the rigid
 limit without any divergence or singularity. On regarding inertial number $I$ as the sole state parameter in a granular material in shear flow, it is implicitly assumed that small
contact deflections due to the finite elastic stiffness of the grains are irrelevant -- this is the rigid limit. 

In the presence of attractive forces between neighboring grains, contacts are endowed with a finite tensile strength 
$F_0$, whence a new dimensionless parameter, $P^*$, a reduced pressure comparing the applied confining stress $P$ (say, the
 controlled normal stress value $\sigma_{\oyy}$ in shear flow) to force scale $F_0$, as 
$P^* = \dfrac{a^2\sigma_{22}}{F_0}$ (similarly a ``cohesion number'' $\eta = 1/P^*$ was defined in Ref.~\cite{RoRoWoNaCh06}). 
Under small $P^*$, cohesion stabilizes loose structures~\cite{KBBW02,GiRoCa07}, which collapse upon increasing $P^*$~\cite{KBBW03,GiRoCa08}. In steady shear flow, 
generalization of rheological laws to the cohesive case involve internal friction coefficient and density, functions of both numbers $I$ and $P^*$ 
or $\eta$~\cite{RRNC08}.

In wet granular materials cohesion arises from capillary forces due to small liquid bridges joining particles touching or in close vicinity to each 
other~\cite{HER05,MiNo06}. The effect of such forces has been investigated in quasistatic deformation~\cite{RYR06,Sou06b,SCND09}, and some of its 
consequences in terms of microstructure were discussed~\cite{RRY06}. In the  \emph{pendular regime} of saturation~\cite{HER05,MiNo06} those 
bridges are small enough and do not merge, so that capillary forces are pairwise additive.  Those attractive forces act as a source of cohesion, and
are also characterized by a small range and some dependence on intergranular distance, as a liquid meniscus might join grains that are not in contact. 

A traditional approach of partially saturated granular materials in geomechanics~\cite{MS05}, which has been investigated in recent
DEM studies~\cite{Catherine,SHNCD09}, is to resort to the concept of \emph{effective stresses}, or stresses such that, 
if applied to the dry material, would produce the same deformation and plastic flow of the granular skeleton as in the ones observed in the wet material. 
Proposed definitions of such an effective stress tensor
in the unsaturated case generalize the Terzaghi principe~\cite{TEPE48} applying to saturated media, 
and  involve a correction of  the average pressure related to saturation and capillary pressure~\cite{Lu2010}.

On the macroscopic scale, the effect of adhesive forces are sometimes described in the quasistatic limit of slow flow by the phenomenological 
Mohr-Coulomb law~\cite{DMWood,BiHi93,AFP13}, 
\be
\sigma_{\oxy} = c+\mu_1^*\sigma_{\oyy} ,
\label{eq:mohrcoulomb}
\ee
characterized by macroscopic cohesion $c$ and internal friction coefficient $\mu_1^*$.

The present paper investigates the constitutive laws applying to wet model granular materials, in the pendular regime, and discusses the influence of 
capillary effects on macroscopic behavior and microstructure. Similarly to refs.~\cite{RoRoWoNaCh06,RRNC08}, the rheology and micromechanical aspects 
are studied for varying $P^*$ and $I$ values (with special emphasis on the quasistatic limit of $I\to 0$). As in dry granular systems and in previous studies on 2D cohesive
materials the material rheology is described in terms of 
apparent friction coefficient (stress ratio) $\mu^*$ and solid fraction $\Phi$ as functions of $I$ and $P^*$, and the applicability of a Mohr-Coulomb relation is  tested. Rheological
and microstructure features as normal stress differences and formation of large clusters bonded by liquid bridges are also investigated.

In the following, we first introduce (Sec.~\ref{sec:model}) the microscopic ingredients of the model material, and report then, 
in Sec.~\ref{sec:homostat}, on the conditions in which homogeneous steady states are observed in shear flows, enabling material constitutive laws to be deduced. Such laws are measured, 
depending on the relevant dimensionless parameters and on some features of the microscopic model, in Sec.~\ref{sec:macro}. Next, in Sec.~\ref{sec:caprheo}, 
we investigate the role of
capillary forces and distant interactions in the material rheology, and revisit the traditional concepts of effective stress and Mohr-Coulomb cohesion. 
Additional studies of microstructural and micromechanical aspects follow: force distributions (Sec.~\ref{sec:fn_unsat}), agglomeration effects (Sec.~\ref{sec:aggl}), structural  anisotropy (Sec.~\ref{sec:aniso_unsat}). The results are discussed and put in perspective in the final, conclusive section~\ref{sec:conc}.  

\section{Model material and simulation setup\label{sec:model}}
We consider a granular assembly composed of $N$ equal-sized spherical beads of diameter $a$, made of a material with Young modulus $E$ and
Poisson ratio $\nu$. The contacts are frictional, satisfying Coulomb's law with friction coefficient $\mu$. 
The granular flow is set by imposing a uniform shear rate $\dot\gamma $ to a rectangular parallelipipedic cell  with edge lengths 
$(L_\alpha)_{1\leq\alpha<3}$.
In order to avoid wall effects periodic boundary conditions are used in all three directions. The periodicity,
in the direction of the flow gradient (direction 2),  is applied with the Lees-Edwards 
procedure~\cite{AT87} and in the two other directions the boundary condition is simple periodic. The 
system size $L_2$ is allowed to fluctuate in order to keep the normal stress $\sigma_{22}$ constant 
equal to a prescribed value $P$ while $L_1$ and $L_2$ are fixed~\cite{PR08a}.

\subsection{Force model}
Elastic and frictional forces are jointly implemented in contacts as in Ref.~\cite{iviso1}, in which a simplified 
Hertz-Mindlin-Deresiewicz force model is used for the elastoplastic contact behaviour.
This model combines the normal Hertz force $F_N$, depending on contact deflection $h$ as
\be
F_N = \frac{\tilde E\sqrt{a}}{3}h^{3/2},
\label{eq:hertz}
\end{equation}
in which we introduced notation $\tilde E = E/(1-\nu^2)$, with a tangential elastic force ${\bf F}_T$, to be 
evaluated incrementally in each time step of the simulation.  The simplification of tangential elasticity 
adopted is that of Ref.~\cite{iviso1}, involving a constant ratio $(2-2\nu)/(2-\nu)$ of tangential ($K_T$) to normal ($K_N$) stiffnesses in contacts, both depending on 
$F_N$, as, from \eqref{eq:hertz}, one has $K_N = \dfrac{dF_N}{dh} \propto F^{1/3}$. Caution should be exercised to avoid spurious 
creation of elastic energy with such laws, and $K_T$ should be suitably rescaled in cases of decreasing normal force and deflection. 
For the details of the elastic model, for the enforcement of the Coulomb condition $\vert\vert {\bf F}_T\vert\vert \le \mu F_N$, and for the objective implementation of
the force law, with due account of all possible motions of a pair of contacting grains, 
the reader is referred to~\cite{iviso1}.

An estimate of  the typical contact deflection under confining stress $P$ defines a dimensionless parameter, stiffness number 
$\kappa$~\cite{RD11},  such that $h/a \propto \kappa^{-1}$. For a Hertzian contact, one may use~\cite{iviso1}
\be
\kappa = (\dfrac{\tilde E}{P})^{2/3}.
\label{eq:siff_num}
\ee
Two values of $\kappa$, 8400 and 39000, used in this study, respectively correspond to glass beads with 
$E=70$~GPa and $\nu=0.3$ under pressures 100~kPa and 10~kPa, as in Ref.~\cite{PR08a}.
Finally, the force model of~\cite{iviso1} which is used here may also comprise a viscous damping term opposing normal relative motion of contacting grains, chosen 
to correspond to a restitution coefficient close to zero in normal collisions. We do not comment this feature, as it was shown~\cite{Dacruz05,PR08a} to have very little influence 
in the slow shear flows of the present study. 

\begin{figure}[h]
  \centering
 \includegraphics[width=6cm]{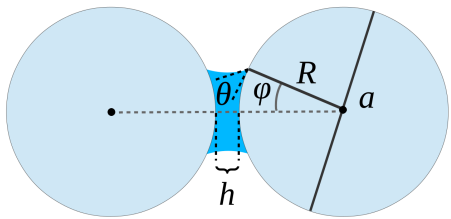}
  \caption{(Color online) A meniscus between two spherical grains of diameter $a=2R$, with distance $h$ between solid surfaces, filling angle $\varphi$, contact angle $\theta$. 
\label{fig:meniscus}}
\end{figure}

The presence of a small amount of an interstitial wetting liquid introduces additional capillary forces, transmitted between 
contacting or neighboring grains by a liquid bridge, or meniscus, as sketched in Fig.~\ref{fig:meniscus}.
We consider a perfectly wetting liquid,  with contact angle $\theta$ equal to zero. As in Ref.~\cite{HER05}, we assume that 
the menisci only form for touching particles, but 
breaks for gaps larger than a certain rupture distance $D_0$, as observed in~\cite{KOH04}. $D_0$ relates 
to meniscus volume $V$ as $D_0\simeq V^{1/3}$~\cite{LiThAd93,WAJS00,PMC00,MAE03}.

For the attractive force between particles separated by distance $h\le D_0$ we adopt the Maugis
approximation~\cite{Maugis87}, which is appropriate for small enough meniscus volume, for its simplicity. 
The maximum attractive force (tensile strength) is reached for contacting particles, 
and equal, according to this model, to $F_0 = \pi a \Gamma$ ($\Gamma$ is the  liquid surface tension) independently of the 
meniscus volume. The 
capillary force varies with the distance $h$ between particle surfaces as
\begin{equation}
   \fcap =
   \begin{cases}
      -F_0 & h \leq 0\\
      -F_0[1-\frac{1}{\sqrt{1+\frac{2V}{\pi a h^2}}}] & 0 < h \leq D_0\\
      0 & h > D_0
   \end{cases}
\label{eq:Maugis}
\end{equation}
$h<0$ corresponds to an elastic deflection of the particles in contact. This formula is a simpler, analytical form of the toroidal approximation 
with the ``gorge method''~\cite{LiThAd93} for the capillary force in a meniscus, which describes the meniscus as limited by circular arcs in a  plane containing the two
sphere centers. 
\begin{figure}[htb]
  \centering
  \includegraphics[width=8cm]{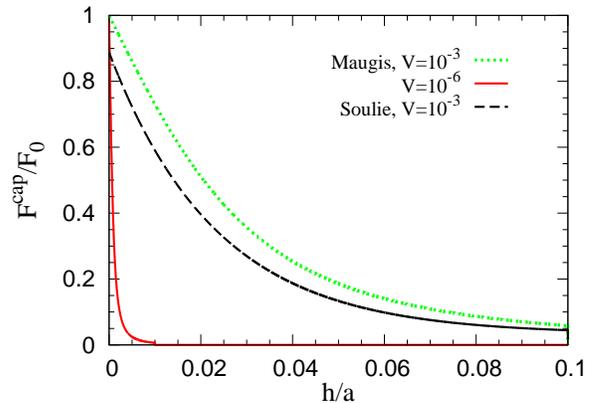}
  \caption{(Color online) Force law $\fcap (h)$, for two different meniscus volumes, according to the Maugis model and to the Souli\'e formula.}
\label{fig:Force_model}
\end{figure}
An alternative form was given by Willett \emph{et al.}~\cite{WAJS00}, while Souli\'e \emph{et al.}~\cite{Sou06a,Sou06b,RR09} proposed a parametrized numerical solution. 
Fig.~\ref{fig:Force_model} compares functions $\fcap(h)$ according to Maugis and to Souli\'e \emph{et al.}. 
Some comparisons between several formulae and experiments are given in~\cite{PMC00}.
\subsection{Saturation range of pendular state\label{sec:satur}}
The morphology of partially saturated granular materials depends on the liquid content~\cite{MiNo06,KUD08}. The present study, like a number of
previous ones~\cite{RYR06,RR09,SCND09}, is restricted to the \emph{pendular state} of low saturations, in which the wetting liquid is confined in bonds or menisci
joining contacting grains.
Liquid saturation $S$ is defined as the ratio of liquid volume $\Omega_l$ 
to interstitial volume $\Omega_v$. It is related to meniscus volume $V$, 
solid fraction $\Phi = 1 - \Omega_v/\Omega$ and wet coordination number $z$ (the average number of liquid bonds on one grain) as:
\begin{equation}
  S = \frac{\Omega_l}{\Omega_v} =  {3z \over \pi} \frac{\Phi}{1-\Phi} \frac{V}{a^3}.
\label{eq:saturation1}
\end{equation}
In our study, we fix the value of meniscus volume $V$. Such a choice does not conserve the total liquid volume, which is proportional to the varying
coordination number $z$ of liquid bonds. Its consequences have to be assessed, and we shall
check that the results are not significantly affected, within the range of investigated material states.

The pendular state to which our model applies is only relevant in some limited saturation range. 
On the one hand, a minimum liquid volume is necessary for menisci to form at contacts, as 
the liquid will first cover the grain surface asperities. 
This minimum saturation $S_{\text{min}}$ for bridges to form might be roughly estimated  upon introducing a roughness scale 
$\delta$,  assuming a layer of thickness $\delta$ covers the surface of the grains, as
\begin{equation}
  S_{\text{min}} =  \frac{6\Phi\delta}{(1-\Phi)a} .
\label{eq:saturation3}
\end{equation}     
For $\Phi = 0.5$ and $\delta\sim 10^{-4} a$ the 
minimum value for saturation is of the order of $10^{-3}$, as observed in experiments~\cite{HER05}. Using~\eqref{eq:saturation1}, and typical values of $z$ (5 or 6) and $\Phi$ (say, 0.6), this sets a lower bound to meniscus volume, of order $10^{-4}a^3$. 
On the other hand, the upper saturation limit for the pendular state corresponds to the merging of the menisci pertaining to the same grain, which, considering a triangle
of spherical grains in mutual contact, happens as soon as the filling angle (see Fig.~\ref{fig:meniscus}) reaches $ \pi/6$.  
The analytical formula  for  $V$~\cite{LiThAd93}, within the toroidal approximation, as a function of $\varphi$ (setting $h=0$, and $\theta=0$), then yields 
$\dfrac{V}{a^3}\simeq 8.10^{-3}$, corresponding, using~\eqref{eq:saturation1}, to a maximum saturation between $0.05$ and $0.1$,  
similar to experimental observations~\cite{HER05,MiNo06}.

\subsection{Choice of parameters~\label{sec:choice}}
Tab.~\ref{tab:list} gives the values of parameters employed in our simulations. While stiffness 
number $\kappa$ and friction coefficient $\mu$ are kept fixed, reduced pressure varies from the dry case $P^* = \infty$ 
down to the lowest value $0.1$, for which cohesive effects are strong, while the 
investigated range of $I$ values allows us to approach the quasistatic limit with some 
accuracy, as well as assess the effects of inertia in faster flows (although rapid, strongly 
agitated flow are not studied here). The meniscus volume is chosen as $V=10^{-3}a^3$ in most simulations.
A few tests are carried out with different values 
(as given within brackets) of the number of particles, the stiffness number and the preset 
meniscus volume. 
\begin{table}[!htb]
\begin{tabular}{c|c}\cline{1-2}
\hline\hline
$\kappa$ & 8400 (occasionally 39000)\\
\hline
$\mu$ & 0.3\\
\hline
$N$& 4000 (8000)\\
\hline
$I$& from $10^{-4}$ to 0.562 by factors of $\sqrt{10}$\\
\hline
$P^*$& 0.1 ; 0.436 ; 1 ; 2 ; 5 ; 10 ; $\infty$\\
\hline
$V/a^3$ & $10^{-3}$ ($10^{-2}$ ; $5\times10^{-3}$ ; $2\times10^{-4}$ ; $10^{-6}$) \\
\hline
\end{tabular}
\normalsize
\caption{\label{tab:list}
List of parameter values: $N$ particles of diameter $a$, interacting with friction 
coefficient $\mu$, forming menisci of volume $V$ at contacts, are subjected to normal 
stress-controlled shear flow for which inertial number $I$, reduced pressure $P^*$ 
(evaluated with normal stress $\sigma_{\oyy}$) and stiffness parameter $\kappa$ take 
values as prescribed. Attractive forces fall to zero at distance $D_0=V^{1/3}$. }
\end{table}
Taking $\Gamma=7.3\times 10^{-2}$~J.m$^{-2}$ for water, and $a=0.1$~mm, $P^*=1$ corresponds
to $\sigma_{\oyy}=(\pi\Gamma/a)\simeq 2.3$~kPa -- the pressure, under gravity, below a granular layer with a thickness of a few tens of centimeters.
\section{Homogeneity and stationarity\label{sec:homostat}}
\subsection{Steady states, macroscopic measurements}
Starting from a dense initial configuration, with solid fraction close to the random close packing value ($\Phi_{\text{RCP}}\simeq 0.64$), we impose a 
constant shear rate $\dot\gamma$ and wait until a steady state is reached before measuring constitutive relations for stresses and solid fraction, which are identified as 
averages over time series. 
Stresses are measured using the standard formula, for all coordinate index pairs $\alpha$, $\beta$
\be
\sigma_{\alpha\beta}=\frac{1}{\Omega}\left[\sum_i m v_i^\alpha v_i^\beta + \sum_{i<j} F_{ij}^\alpha r_{ij}^\beta\right],
\label{eq:stress}
\ee
involving a kinetic contribution with a sum over grains $i$, of velocities $v_i$ and a sum over all pairs with center-to center vector $r_{ij}$, interacting with force $F_{ij}$, $\Omega$
denoting the sample volume.

\begin{figure}[h]
   \centering
   \includegraphics[width=.9\linewidth]{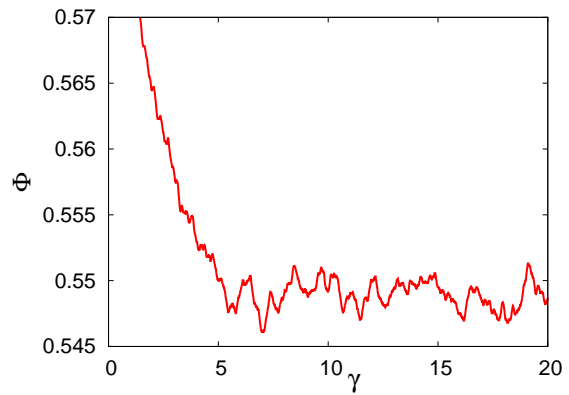}
   \caption{(Color online) Solid fraction $\Phi$ versus shear strain $\gamma$. Time series is obtained with 
 	$P^*=1$, $I=0.1$ and $N=4000$ when the rupture distance is $D_0=0.1$.}
\label{fig:phi_g}
\end{figure}

\begin{figure}[h]
   \centering
   \includegraphics[width=1.\linewidth]{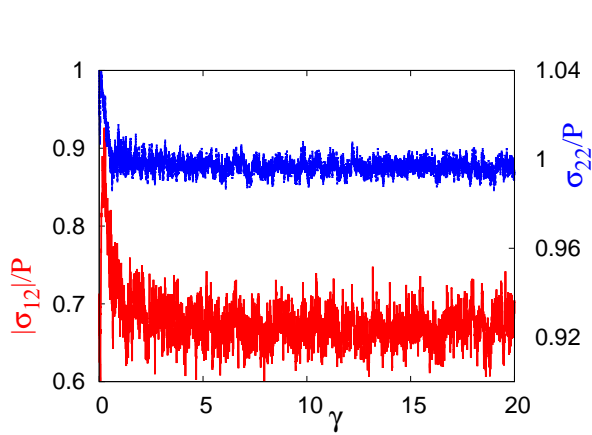}
  \caption{(Color online) Shear stress $|\sigma_{\oxy}|$ (lower curve, red, left axis) and normal stress $\sigma_{\oyy}$ (upper curve, blue, right axis) 
 	versus shear strain $\gamma$. Note the different scales on left and right axes. Time series  obtained with 
 	$P^*=1$, $I=0.1$ and $N=4000$ when the rupture distance is $D_0=0.1$.}
\label{fig:sxy_g}
\end{figure}

The evolution of solid fraction $\Phi$ with strain $\gamma$ is shown in Fig.~\ref{fig:phi_g}:  $\Phi$ decreases until it approaches its steady state value 
for $\gamma \ge 5$ in this case.  Fig.~\ref{fig:sxy_g} shows the evolutions of $\sigma_{\oyy}$ and $\sigma_{\oxy}$ with 
$\gamma$. (Note that $\sigma_{\oxy}$ is negative with our sign convention). 
We thus check that normal stress $\sigma_{\oyy}$ is well controlled since its fluctuations about its  prescribed value $P$ are very small.
Shear stress $\sigma_{\oxy}$ exhibits a fast increase and and overshoot at small strain and then decreases, approaching its steady state value, after a few strain units, 
over a strain interval (a few units) similar to the one corresponding to the transient evolution of $\Phi$.
Stresses and solid fraction fluctuate  in the steady state, and a careful evaluation of measurement errors on their time averages is required (especially for shear stress, 
for which fluctuations levels reaching  about $20\%$ of the mean value are apparent in the example of Fig.~\ref{fig:sxy_g}). We use the 
blocking technique of Ref.~\cite{FP89} to estimate error bars on averages over finite time series.
\subsection{Shear localization\label {sec:loc}}
\subsubsection{Velocity profile}
\begin{figure}[htb]
  \centering
  (a) \includegraphics[width=1.\linewidth]{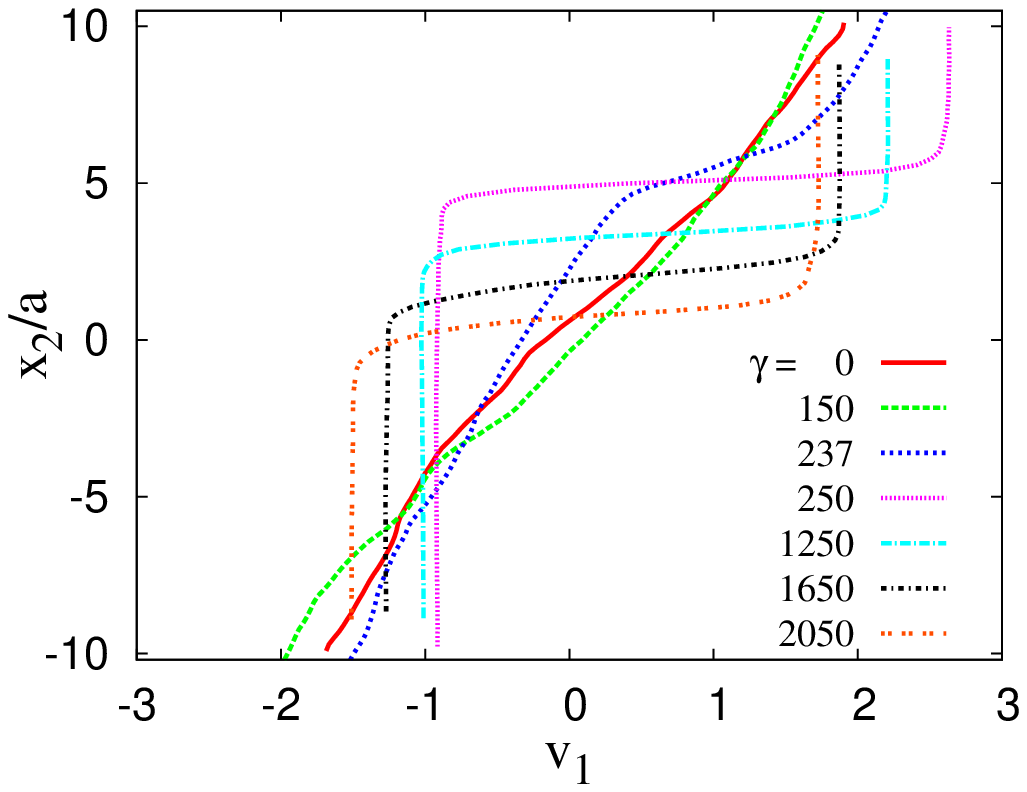} \\
  (b) \includegraphics[width=1.\linewidth]{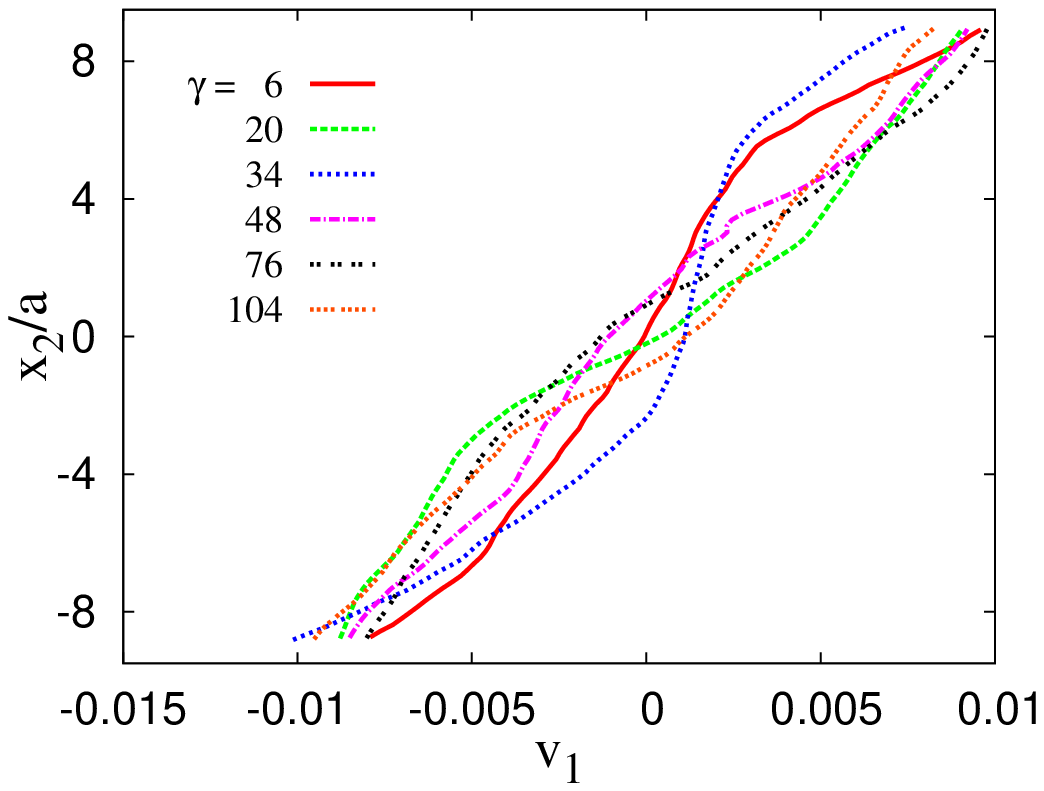}
  \caption{(Color online) Velocity profile for $P^* = 0.1$, $I = 0.178$ (a) and for $P^* = 0.436$ and 
  $I = 10^{-3}$ (b) at different shear strain values.}
\label{fig:velprof1_2}
\end{figure}

Instantaneous velocity profiles $v^s(\equiv\langle v_{\ox}(x_2)\rangle_s)$ are computed on averaging 
particle velocities along the mean flow direction within slices of thickness $0.01L_{\oy}$. 
We observe a strong localization of the flow for the smallest value of the reduced pressure, $P^*=0.1$,
for both  slow and  flows. As represented in Fig.~\ref{fig:velprof1_2}a, the velocity gradient, initially uniform,  
gradually   concentrates within a shear band of thickness $H\lesssim 3a$
which may move vertically but persists for all values of strain $\gamma>250$. Localization tendencies
in slow flow of dry granular materials are sometimes reported~\cite{AhSp02,Dacruz05,PR08a}, although, for uniform strain rates, 
usually not observed as an enduring, systematic phenomenon.
In the present study persistent localization profiles are also detected in nearly quasistatic flows, with a shear band thickness between $5a$ to $10a$. 
However, for the intermediate values of the inertial number ($10^{-2}\lesssim I\lesssim 10^{-1}$) this 
effect diminishes and the strongly localized profiles are less frequent.

For all $P^*\ge 0.436$, localization is not frequently observed and temporary, even in the 
quasistatic limit. The velocity profiles for $P^*=0.436$ and $I=10^{-3}$, as represented in 
Fig.~\ref{fig:velprof1_2}b, are nearly linear and on average the flow is homogeneous.

\subsubsection{Local solid fraction}
\begin{figure}[h]
  \centering
  (a) \includegraphics[width=1.\linewidth]{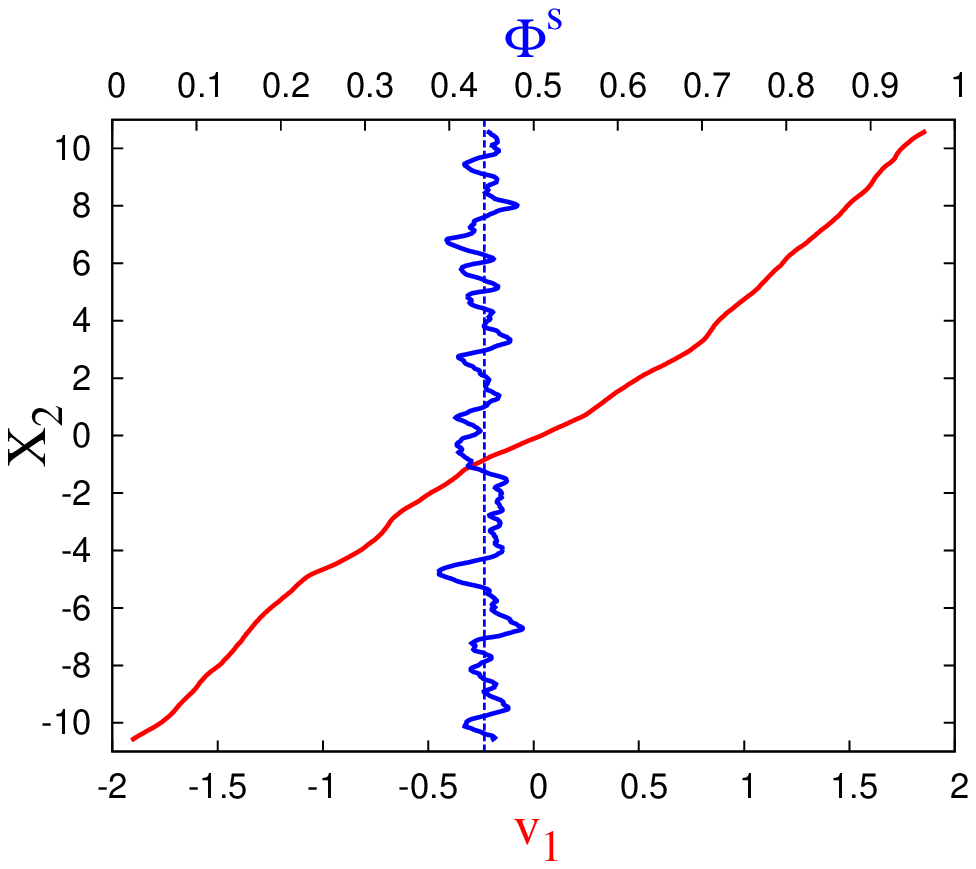}\\
  (b) \includegraphics[width=1.\linewidth]{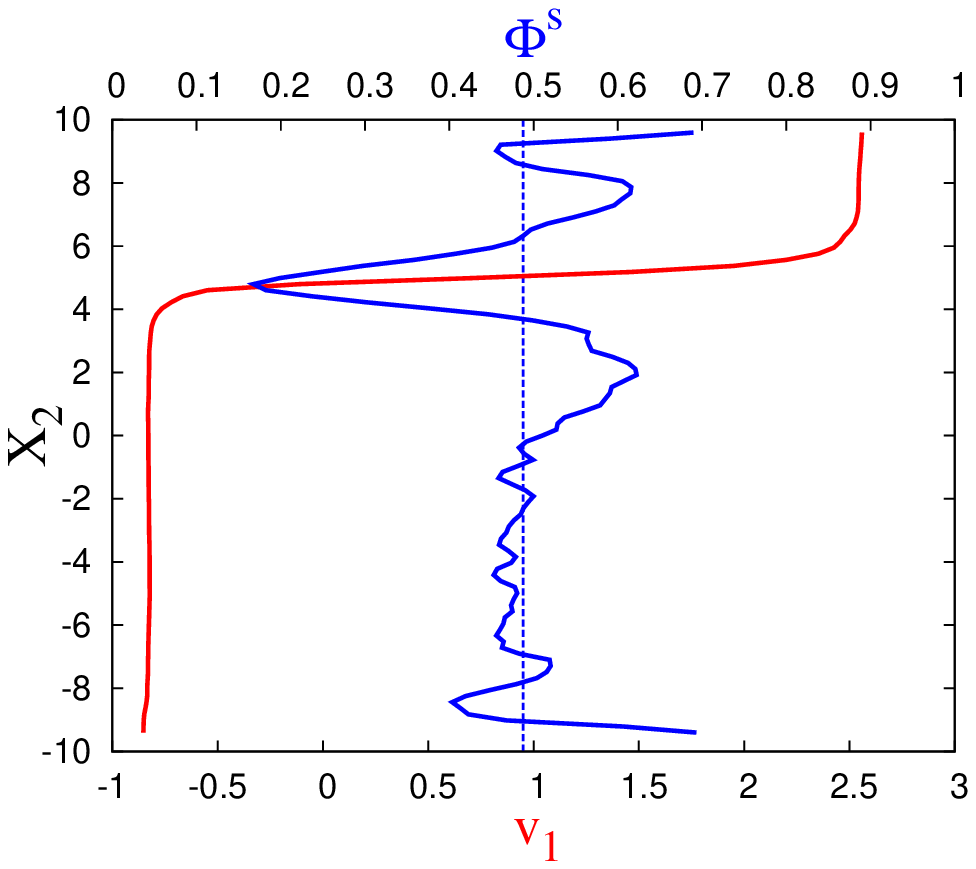}
  \caption{(Color online) Velocity profiles (lower axis in red) and local densities of grains (upper axis in blue) 
	  for two configuration samples with shear strain $\gamma=20$ (a) and $\gamma=352$ (b) 
	  when $P^*=0.1$ and $I=0.178$. Average solid fraction $\langle\Phi^s\rangle$, is shown 
	  with a vertical dashed line in blue, with a value of 0.44 in (a) and 0.49 in (b).}
\label{fig:phi_s_g017}
\end{figure}

Similarly we can calculate a solid fraction profile, on averaging the solid contents of slices 
orthogonal to the velocity gradient (splitting the volume of one grain between different 
slices if necessary). Fig.~\ref{fig:phi_s_g017} shows the velocity ($v^s$) and solid fraction ($\Phi^s$) profiles
for two different values of shear strain, $\gamma=1$ and $\gamma=352$, which belong to the simulation 
of Fig.~\ref{fig:velprof1_2}a. We see that in the homogeneous flow the distribution of mass in the 
system is almost uniform, but when the localization occurs $\Phi^s$ strongly decreases 
within the shear bands, to a value below 0.2. It slightly increases outside the shear band, especially 
in its vicinity. Slighter decrease of density within thicker shear bands in quasistatic flow is 
observed. For instance, when $I=10^{-3}$, $\Phi^s$  decreases  from 0.47 to about 0.4
inside the sheared zone of thickness $H\approx 7a$. 

\subsubsection{Deviation from linear profile}
\begin{figure}[htb]
  \centering
  \includegraphics[width=0.9\columnwidth]{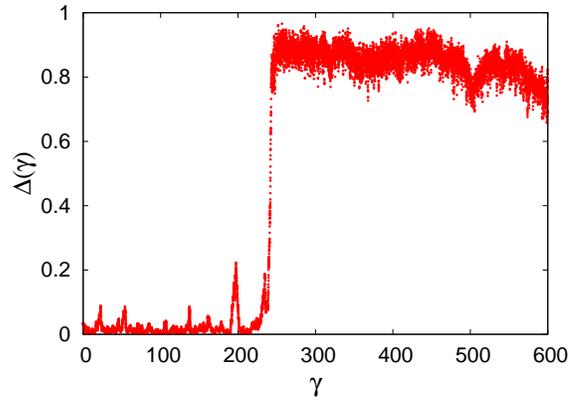}
  \caption{(Color online) Deviation from linear profile, $\Delta(\gamma)$, versus strain $\gamma$ 
       for $P^* = 0.1$, $I = 0.178$.}
\label{fig:dfl1_2}
\end{figure}

The deviation from the linear profile is characterized by parameter $\Delta$:
\begin{equation}
  \Delta (t) = \frac{12}{L_{\oy}^3 \dot{\gamma}^2} \int\limits_{-L_{\oy}/2}^{L_{\oy}/2} (v_{\ox}(\xoj) - \dot{\gamma} \xoj)^2 d\xoj.
  \label{eq:dfl_velprof1}
\end{equation}
The normalization by $\dfrac{L_{\oy}^2}{12}$ ensures a maximum value $\Delta=1$ in the case 
of a perfect localization within a plane, as if two solid blocks were sliding on each other.
As defined in Eq.~\eqref{eq:dfl_velprof1}, parameter $\Delta$ is not 
affected by a vertical shift in the velocity profile, due to the periodic boundary 
condition in direction $x_2$. 
 If the strain rate is homogeneous within a shear band of 
thickness $H$ and vanishes outside, one observes
\be
  \Delta = (1-{H\over L_{\oy}})^2
\label{eq:D_HL}
\ee
Fig.~\ref{fig:dfl1_2} is the plot of $\Delta$ as a function of strain $\gamma$ 
corresponding to the same simulation as in Fig.~\ref{fig:velprof1_2}a. It initially shows small fluctuations 
near zero, and suddenly increases near $\gamma=250$ when the velocity gradient localizes in a shear band.
\subsubsection{Occurence of shear banding}
Large values $\Delta>0.8$ for  $P^*=0.1$ in the faster flows ($I\ge 0.17$) indicate strong 
localization in this range. At $I=0.1$ $\Delta$ drops down to small values, typically below 0.1, 
but in the quasistatic limit it increases again: at $I=10^{-3}$ it  mainly fluctuates between 0.4 and 0.8.
For $P^*\geq 0.436$ the shear rate is much more homogeneous. $\Delta$ almost vanishes in faster flows,
increases somewhat in the quasistatic limit, but rarely exceeds 0.2, even for the smallest inertial numbers.

Simulations carried out with a larger stiffness number ($\kappa=39000$), 
for the two smallest values of $P^*$ and for all values of $I$ in Tab.~\ref{tab:list},
 do not record any significant influence of $\kappa$ on the homogeneity of the flow. 
 The influence of the sample size is studied by simulating some 
samples with height $L_{\oy}$ twice as large as in the standard sample, containing  8000 grains, with $P^*=0.436$
and different values of $I$. The size dependence in formula~\ref{eq:D_HL} implies then larger values of $\Delta$, 
should the shear strain tend to localize, temporarily or permanently, 
 within a region of fixed thickness. In our tall, 8000 grain systems, as the quasistatic limit is approached, 
 $\Delta$  reaches peak values above 0.4 but continuously 
evolves and no persistent localization pattern is detected.

Consequently, our results reveal a strong localization tendency at $P^*=0.1$ for both small  (below 0.03) and large  (above 0.3) values
of the inertial number. We performed some measurements at $P^*=0.1$ for intermediate values of $I$, over strain intervals for which values of 
inhomogeneity parameter $\Delta$ averaged below $0.1$, as in the first part of the graph of Fig.\ref{fig:dfl1_2}.
We limited our systematic studies to homogeneous flows, for larger values of $P^*$, for which no evidence
of enduring localization effects is observed. 

A systematic fluid depletion in shear bands was reported in~\cite{MKOH12} -- this requires a model for liquid migration between menisci, which 
we did not introduce in the present study. We limit our results to the issue of whether shear banding occurs, and focus on homogeneous flows.
Shear banding and inhomogeneous flows certainly deserve more detailed studies, which we plan to pursue in another publication.

\section{Macroscopic behaviour and constitutive relations\label{sec:macro}}
Focussing on homogeneous steady states (excluding too small $P^*$ values, from Sec.~\ref{sec:loc}) we now deduce 
macroscopic constitutive relations from the simulations.

\subsection{Shear stress and solid fraction}
\begin{figure}[htb]
  (a) \includegraphics[width=1.\linewidth]{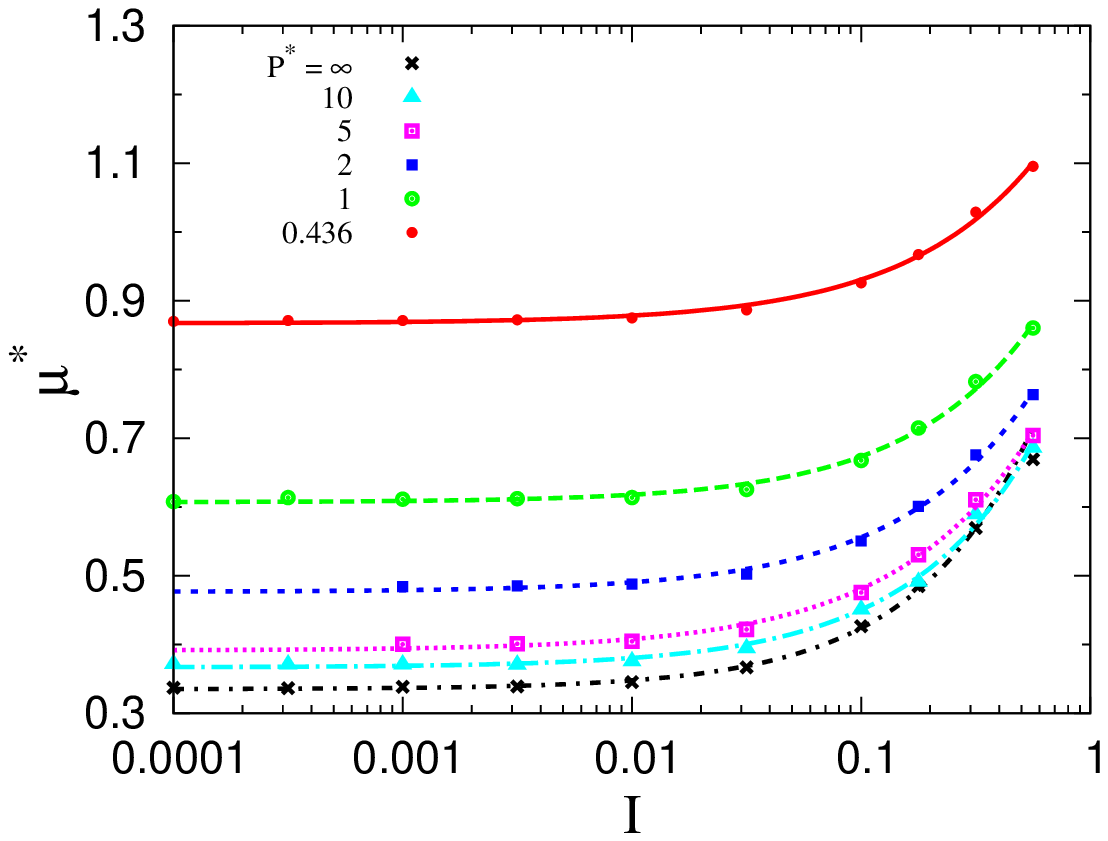}\\
  (b) \includegraphics[width=1.\linewidth]{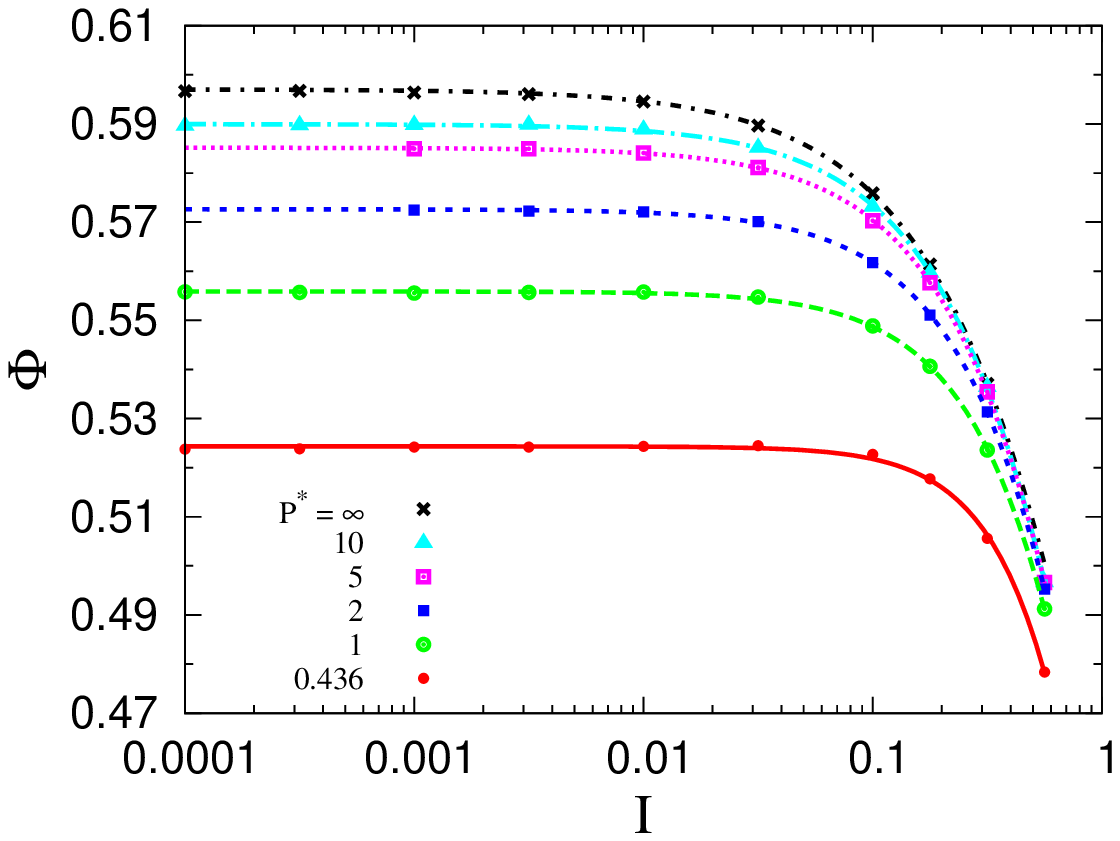}
  \caption{(Color online) Macroscopic friction coefficient $\mu^*$ (a) and solid fraction $\Phi$ (b) versus inertial number 
	    $I$ for different values of reduced pressure $P^*$ ($D_0 = 0.1 a= V^{1/3}$).}
\label{fig:mu_phi}
\end{figure}

Friction coefficient $\mu^*$ and solid fraction $\Phi$, depending on $I$ for various $P^*$ values, are shown in 
Fig.~\ref{fig:mu_phi} for the parameter choice adopted in most simulations.
We fit the following power law functions to those data,  denoting as $\mu^*_0$ and $\Phi_0$ 
the quasistatic limits (critical state) values of the macroscopic friction coefficient and of the solid fraction:
\be
  \begin{cases}
    \mu^* = \mu^*_0 +cI^{\alpha} \\
    \Phi^{-1} = \Phi^{-1}_0 +eI^{\nu}
  \end{cases}
\label{eq:func_I}
\ee
Tabs.~\ref{tab:fit_mu} and~\ref{tab:fit_phi} give the values of the fitting parameters introduced in Eqs.~\ref{eq:func_I}.
While the increase of $\mu*$ and the decrease of $\Phi$ as functions of $I$ are familiar trends, 
similar to observations made with dry grains~\cite{Dacruz05,Hatano07,PR08a,AzRa2014},
some other features are remarkable. The quasistatic limit is quite nearly approached for $I\le 0.01$, and is strongly influenced by capillary 
forces. 
Internal friction coefficient $\mu*$, compared to the dry, 
cohesionless value (0.335$\pm 0.001$), already shows a notable increase at $P^*=10$, reaching values as high as $0.6$ 
for $P^*=1$ (\emph{i.e.}, as cohesive and confining forces are of the same order), and nearly $0.9$ for $P^*=0.436$, about 2.3 times the cohesionless value.
Our partial results for $P^*=0.1$, measured in reasonably homogeneous flows, indicate $\mu^*\simeq 1.6$ for $I=10^{-2}$. 
Meanwhile, the material becomes looser, with $\Phi$ reaching values that cannot be observed without cohesion in quasistatic conditions. 
\begin{table}[!htb]
  \centering
  \begin{tabular*}{0.9\linewidth}{@{\extracolsep{\fill}} l c c c }
      \hline\hline
      $P^*$ 	& $\mu^*_0$ 		& $\alpha$ 		& $c$ \T \\
      \hline
      0.436	& $0.867\pm 0.003$ 	& $0.76\pm 0.04$ 	& $0.36\pm 0.01$ \T \\
      1		& $0.607\pm 0.003$ 	& $0.79\pm 0.05$ 	& $0.41\pm 0.02$ \\
      2		& $0.477\pm 0.005$ 	& $0.76\pm 0.05$	& $0.45\pm 0.02$ \\
      5		& $0.391\pm 0.006$ 	& $0.73\pm 0.05$	& $0.49\pm 0.02$ \\
      10	& $0.367\pm 0.004$ 	& $0.79\pm 0.04$ 	& $0.51\pm 0.02$ \\
      $\infty$	& $0.335\pm 0.001$ 	& $0.84\pm 0.02$	& $0.62\pm 0.02$ \\
      \hline\hline
  \end{tabular*}
  \normalsize
  \caption{Parameters of the fit of function  $\mu^*(I)$ by Eq.~\ref{eq:func_I}, for different values of $P^*$.}
\label{tab:fit_mu}
\end{table}
  
\begin{table}[!htb]
  \centering
  \begin{tabular*}{0.9\linewidth}{@{\extracolsep{\fill} } l c c c }
      \hline\hline
      $P^*$ 	& $\Phi_0$ 			& $\nu$ 		& $e$ \T\\
      \hline
      0.4360	& $0.5243\pm 2.10^{-4}$ 	& $1.73\pm 0.05$	& $0.497\pm 0.017$ \T\\
      1		& $0.5559\pm 10^{-4}$ 		& $1.34\pm 0.012$	& $0.512\pm 0.005$ \\
      2		& $0.5726\pm 10^{-4}$ 		& $1.21\pm 0.01$	& $0.547\pm 0.003$ \\
      5		& $0.5851\pm 10^{-4}$		& $1.12\pm 0.01$	& $0.580\pm 0.003$ \\
      10	& $0.5900\pm 10^{-4}$		& $1.09\pm 0.01$	& $0.594\pm 0.004$ \\
      $\infty$	& $0.5970\pm 10^{-4}$	& $0.96\pm 0.015$	& $0.562\pm 0.008$ \\
      \hline\hline
  \end{tabular*}
  \normalsize
  \caption{Parameters of the fit  of function  $\Phi(I)$ by Eq.~\ref{eq:func_I}, for different values of $P^*$.}
  \label{tab:fit_phi}
\end{table}

Such a strong influence of cohesive (capillary) forces contrasts with the results of Refs.~\cite{RoRoWoNaCh06,RRNC08}, in which similar deviations 
between cohesionless and cohesive systems are not observed until $P^*$ decreases to much lower values, of order $0.01$. 
Such 2D results were however obtained with a different attractive 
force law, of vanishing  range beyond contact. The origins of the strong rheological effects of capillary forces are investigated in the following.
\subsection{Normal stress differences}
The first and the second normal stress differences are defined as 
\be
  \begin{cases}
    N_1 &= \sigma_{\oxx}-\sigma_{\oyy} \\
    N_2 &= \sigma_{\oyy}-\sigma_{\ozz}
  \end{cases}
\label{eq:nsd_12}
\ee
Note that those definitions coincide with the one used in complex fluid or suspension rheology~\cite{GuJe12}, but that we use the opposite sign convention for normal stresses.
Signs of $N_1$ and $N_2$ should thus be reversed for comparisons to this literature. 

\begin{figure}[htb]
  \centering
  (a)  \includegraphics[width=1.\linewidth]{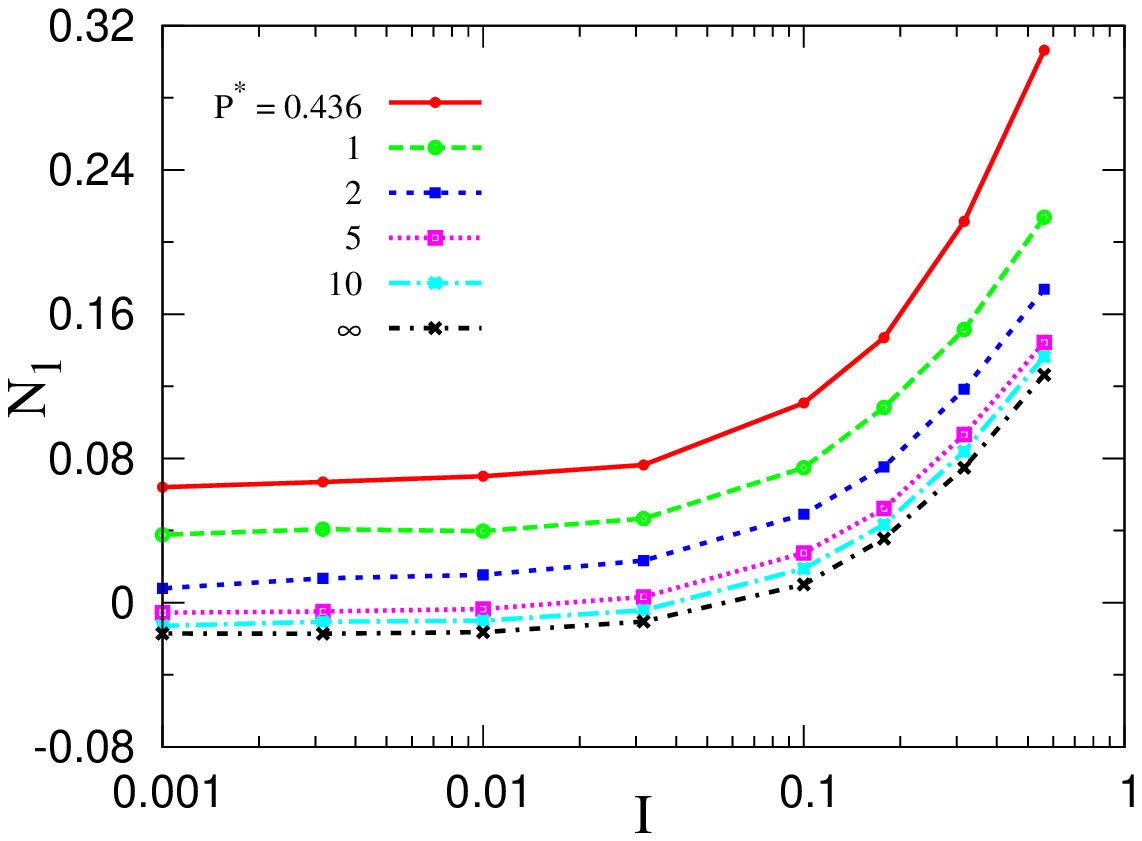}
  (b)  \includegraphics[width=1.\linewidth]{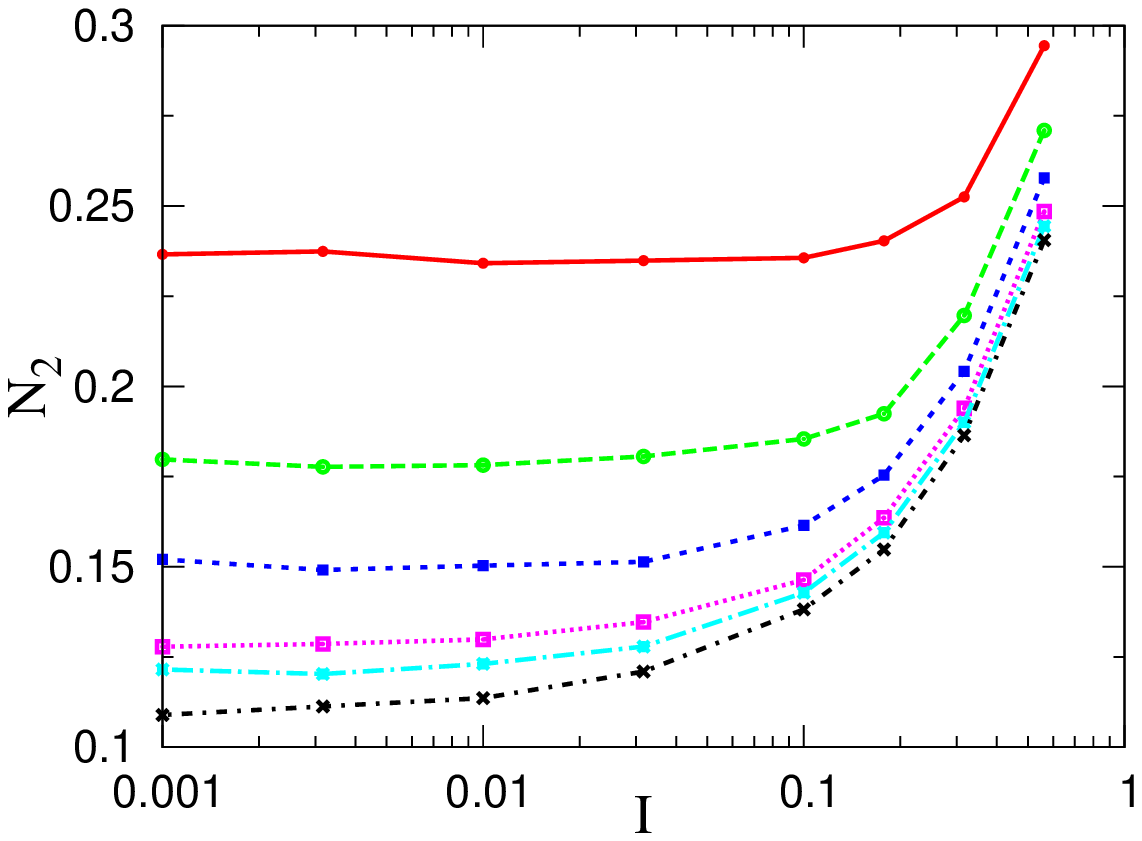}
  \caption{(Color online) First (a) and second (b) normal stress differences as functions of $I$ for different $P^*$ values. 
  The same symbols and colour codes apply to both figures.}
\label{fig:str_diff}
\end{figure}

Most often, considering dense flows of dry granular materials, those differences, deemed small, are ignored or neglected~\cite{AFP13}. We find it worthwhile to record their
values nevertheless, since, as shown
in Fig.~\ref{fig:str_diff}, where $N_1$ and $N_2$ are plotted versus $I$ for different values of $P^*$, they are strongly influenced by capillary forces.
The first normal stress difference is very small in the
quasistatic limit and for large values of the reduced pressure. It increases with $I$ and for decreasing values of $P^*$, 
going through a transition from small  
negative values to positive values near $I=0.03$, for $P^*\ge 2$. $N_1$ variations with $I$ are nearly parallel 
for different $P^*$ values. 
The second normal stress difference $N_2$ also increases for faster flows and for decreasing reduced pressure $P^*$. In the 
quasistatic limit, it is considerably larger then $N_1$. 
\subsection{Sensitivity to capillary force model and saturation}
\subsubsection{Capillary force model}
We tested the effect of the capillary force model
by replacing the Maugis approximation, Eq.~\ref{eq:Maugis}, with the more accurate parametrized capillary force law 
proposed by Souli\'e \emph{et al.}~\cite{Sou06b,Gras09}, for $V=10^{-3}a^3$. Although the difference in the force models is appreciable on a plot 
of $\fcap$ versus $h$ (with the Souli\'e force about 10\% smaller at contact, see Fig.~\ref{fig:Force_model}), the macroscopic results are very close: 
the difference in stress ratio $\mu^*$ and solid fraction $\Phi$ increases with $I$ but does not exceed 2\%. 
\subsubsection{Meniscus volume and force range}
\begin{figure}[h]
  \centering
  (a) \includegraphics[width=1.\linewidth]{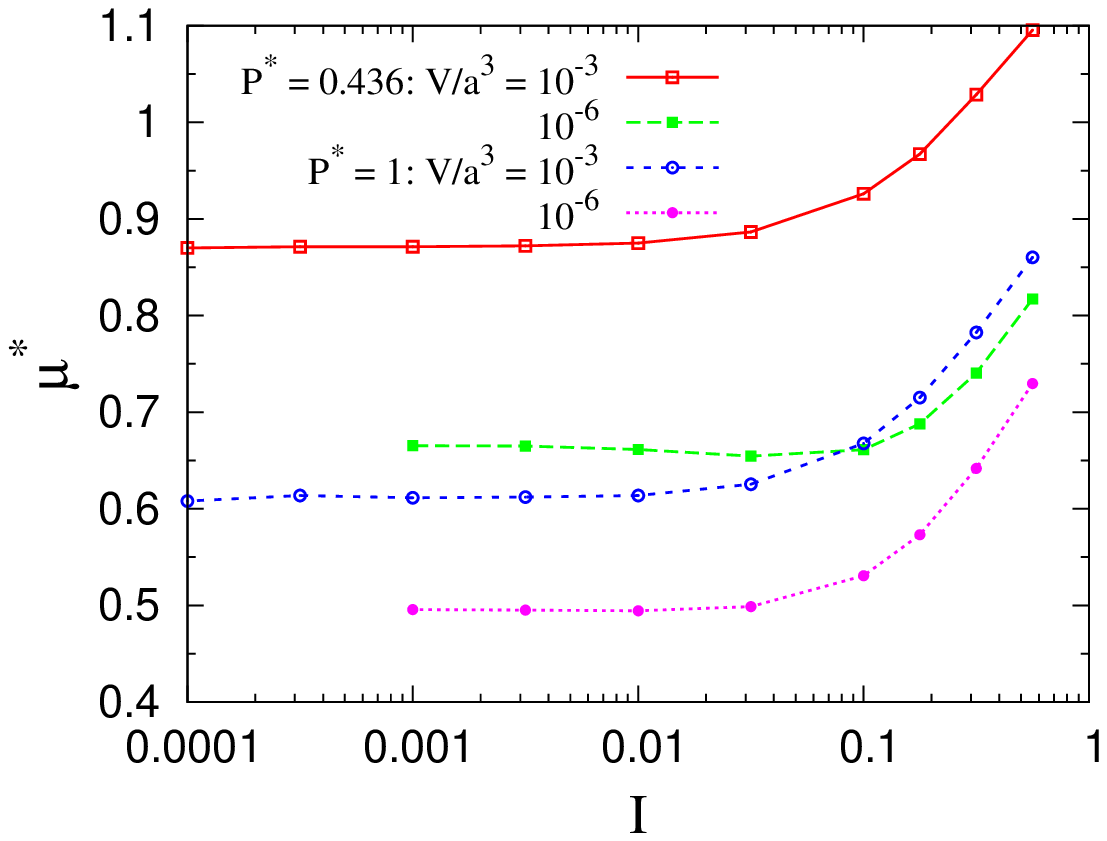} \\
  (b) \includegraphics[width=1.\linewidth]{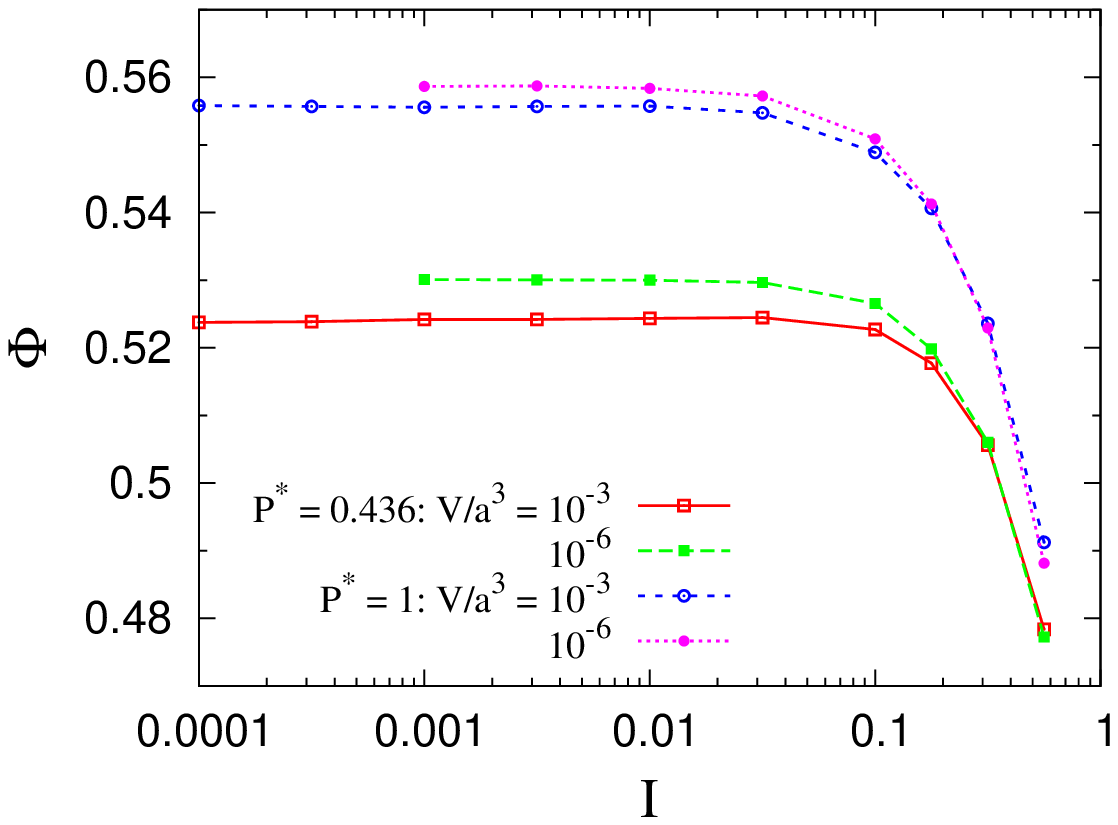}
  \caption{(Color online) Macroscopic friction coefficient $\mu^*$ (a) and solid fraction $\Phi$ (b) versus 
	   inertial number $I$ for different values of $P^*$ and meniscus volume $V$ (with 
	   $D_0=V^{1/3}$).}
\label{fig:vol_mu_phi}
\end{figure}

Changing the meniscus volume amounts to changing the distance at which the attractive force vanishes, rupture distance 
$D_0=V^{1/3}$, as well as the gap dependence of the capillary force $\fcap (h)$ (Fig.~\ref{fig:Force_model}).
Fig.~\ref{fig:vol_mu_phi} shows internal friction coefficient  $\mu^*(I)$  to be strongly sensitive to meniscus volume 
for the lowest $P^*$ values. For a meniscus volume of $10^{-6}a^3$, as compared to the standard value  $10^{-3}a^3$, $\mu^*$ decreases by about $20\%$.
Actually, for such a small meniscus volume, the decay of the attractive force (Fig.~\ref{fig:Force_model}) is so fast that, as we checked,
results are hardly changed on setting the meniscus rupture distance to zero. The effect on the solid fraction $\Phi$ remains small.
To explore the rheological properties throughout the pendular regime, 
we varied the meniscus volume, and recorded the solid fraction and the friction coefficient in the quasistatic limit for the smallest studied $P^*$ value, 
as indicated in Tab.~\ref{tab:sat_eff}, thus fully covering the corresponding saturation range (see Sec.~\ref{sec:satur}).
Saturation $S$, by relation~\eqref{eq:saturation1}, is related to the wet coordination number, $z$, whose values are also provided in the table. 
While the change in solid fraction  does not exceed $0.01$, the variation of the macroscopic friction coefficient is about $20\%$ in the pendular regime 
(up to $50\%$ upon extending the numerical study  to unrealistically small menisci, $V=10^{-6}a^3$). 
\begin{table}[!htb]
 \centering
    \begin{tabular*}{0.9\linewidth}{@{\extracolsep{\fill} } l c c c c }
      \hline\hline
      $V/a^3$ 		& $S_w$ 			& $z$ 		& $\Phi$ 	& $\mu^*$ \\
      \hline
      $10^{-2}$		& $7.137\times 10^{-2}$ 	& $6.863$	& $0.520$	& $1.071$ \\
      $5\times 10^{-3}$	& $3.418\times 10^{-2}$ 	& $6.556$	& $0.522$	& $1.003$ \\
      $10^{-3}$		& $6.305\times 10^{-3}$ 	& $5.970$	& $0.524$	& $0.875$ \\
      $2\times 10^{-4}$	& $1.075\times 10^{-3}$		& $5.534$	& $0.525$	& $0.787$ \\
      $10^{-6}$		& $5.539\times 10^{-6}$		& $4.836$	& $0.530$	& $0.661$ \\
      \hline
    \end{tabular*}
    \normalsize
    \caption{(Color online) Effect of meniscus volume or saturation level on different 
    parameters for $I=10^{-2}$ and $P^*=0.436$.}
\label{tab:sat_eff}
\end{table}
We therefore predict a moderate variation of rheological properties within the simulated pendular regime of the partially saturated granular assembly.
Returning to the basic assumptions of our model, one of its drawbacks is that it ignores liquid volume conservation. Within the granular sample, the total
liquid volume is proportional to coordination number $z$. As $z$ varies with $I$, we should in principle correct the meniscus volume to maintain 
a constant product $zV$ for different shear rates. However, the $V$ dependence of macroscopic properties is so slow ($\mu^*$ varies by 20\% as $V$ is multiplied by $20$)
that the resulting correction on $V$ (as $z$ changes, typically, from 6 to 4 at most) should not exceed 2\%. 
\subsubsection{Hydraulic hysteresis}
\begin{figure}[h!]
  \centering
  (a) \includegraphics[width=1.\linewidth]{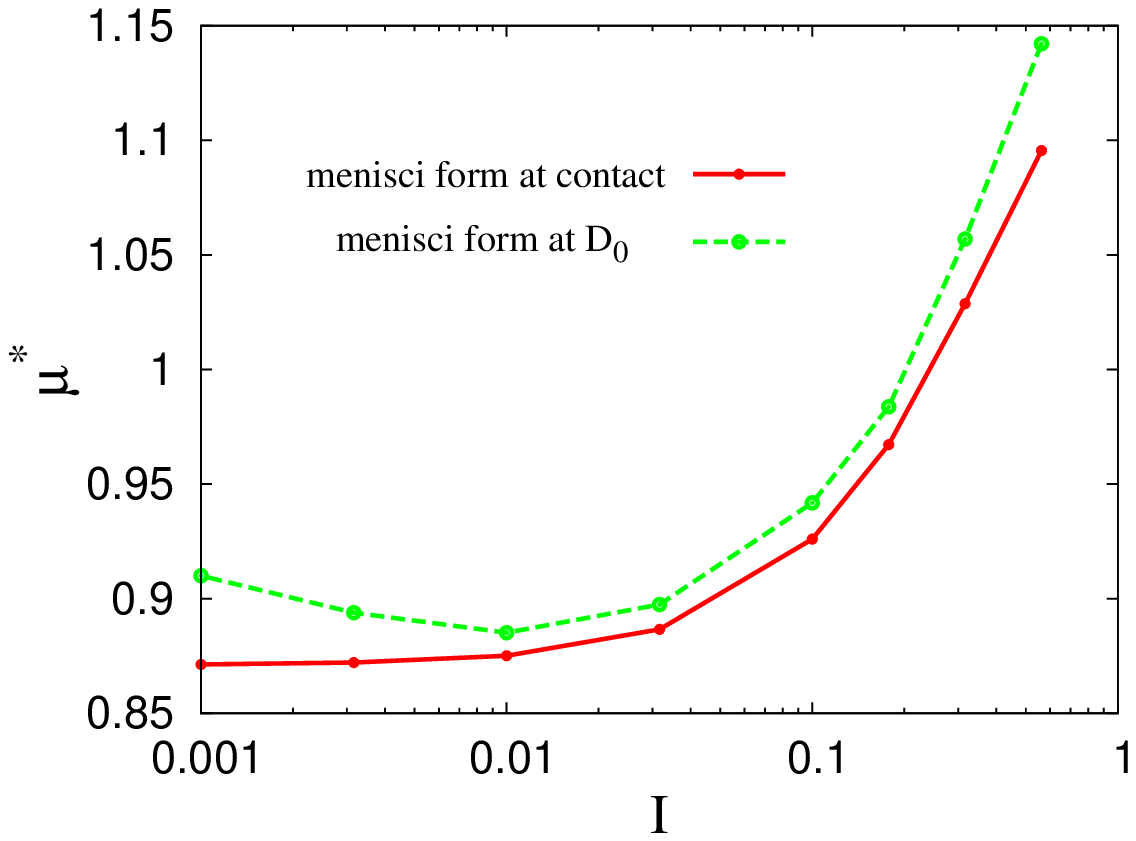} \\
  (b) \includegraphics[width=1.\linewidth]{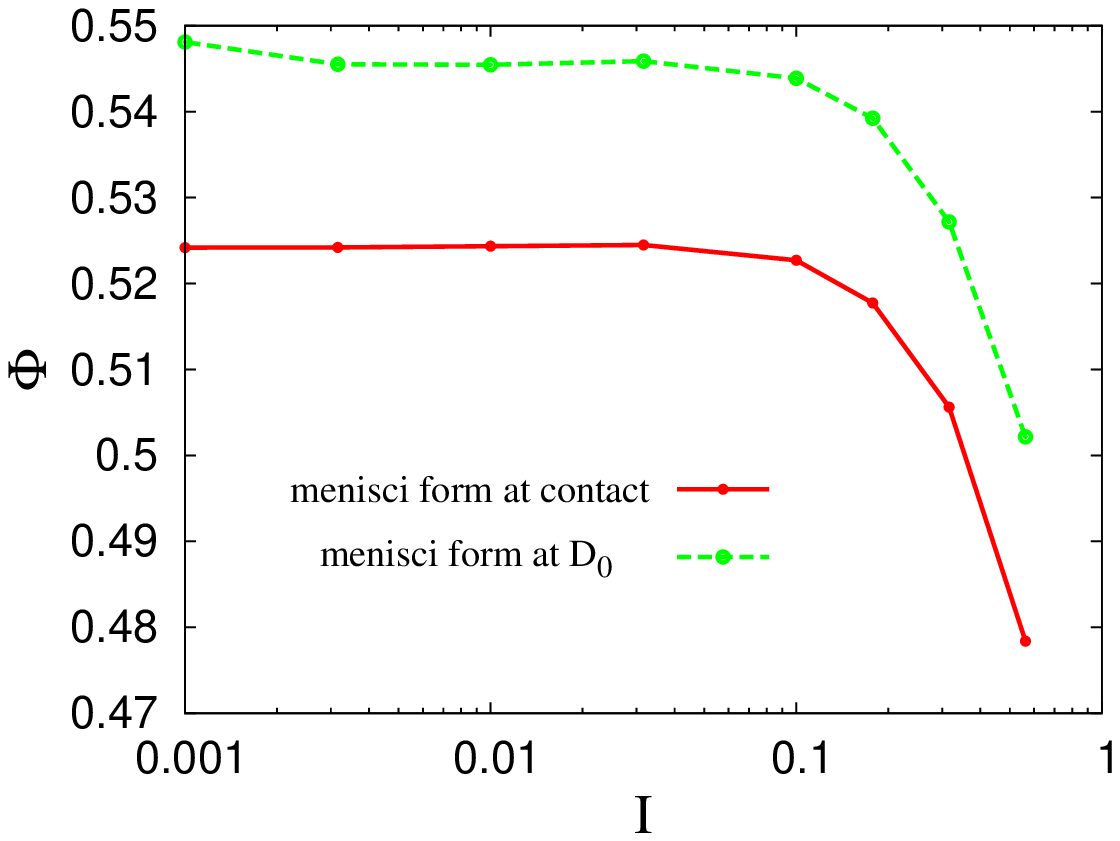}
  \caption{(Color online) Macroscopic friction coefficient $\mu^*$ (a) and solid fraction $\Phi$ (b) versus 
	   inertial number $I$ for $P^*=0.436$ and $D_0=0.1$, with and without capillary hysteresis.}
\label{fig:mu_phi_hyst}
\end{figure}

Another feature of meniscus model the role of which should be explored is the hysteresis of the attractive force, which appears at contact, and vanishes at distance $D_0$.
Given the results of Tab.~\ref{tab:frac_lb}, one may expect a strongly enhanced influence of distant interactions (as reported, e.g., in Ref.~\cite{ShGr08}). 
Fig.~\ref{fig:mu_phi_hyst} compares internal friction and solid fraction, for different values of $P^*$ and $I$, in the standard, hysteretic model, and 
without the capillary force hysteresis, assuming menisci to appear as soon as non-contacting grains approach below distance $D_0$. 
With the new rule of meniscus formation, $\Phi$ strongly increases, especially for small values of $I$. 
The internal friction $\mu^*$, for $I\simeq 0.1$, is close to the standard case, but larger values are obtained as $I$ decreases. 
Even for the smallest values of $I$ investigated ($I=0.001$), the material properties still depend on shear rate
and no proper critical state appears to be approached in our simulations. 
The decrease of $\mu^*$ as a function of $I$ in interval $0.001\le I\le 0.01$ should trigger shear-banding instabilities, 
as discussed in~\cite{ZShojaaee,SRCW12}. A slightly decreasing trend of $\mu^*$ versus $I$ was also apparent in Fig.\ref{fig:vol_mu_phi}, for very small $D_0$. 
For the standard value $V=10^{-3}a^3$ adopted in this study (as one for which laboratory observations should be possible), 
the friction coefficient does increase with $I$, albeit slower and slower as $P^*$ decreases (see coefficient $c$ in Tab.\ref{tab:fit_mu}). 
The stabilizing effect of this growing variation is weaker as cohesion gets stronger, which is consistent with the systematic shear banding behaviour at $P^*=0.1$, 
and might be jeopardized on tampering with the capillary force model. 

\section{Rheological effect of capillary forces\label{sec:caprheo}}
We now seek to explain the strong influence of capillary forces on the macroscopic material rheology. The roles of different interactions, in the force network and in the stresses
are  investigated. We first collect information on coordination numbers and neighbor distances (Sec.\ref{sec:coord}). Simple relations to average forces are recalled
in Sec.~\ref{sec:forcemoy}.
We split the stresses into several contributions, in order to appreciate the importance of different types of forces. This decomposition (Sec.~\ref{sec:split}) suggests 
an attempt to relate the rheology of wet grains to that of dry ones, in terms of some "effective pressure" approach, in the quasistatic limit, which we present in Sec.~\ref{sec:discu}.
\subsection{Coordination numbers and near neighbor distances\label{sec:coord}}
\begin{figure}[htb]
  \centering
  (a) \includegraphics[width=1.\linewidth]{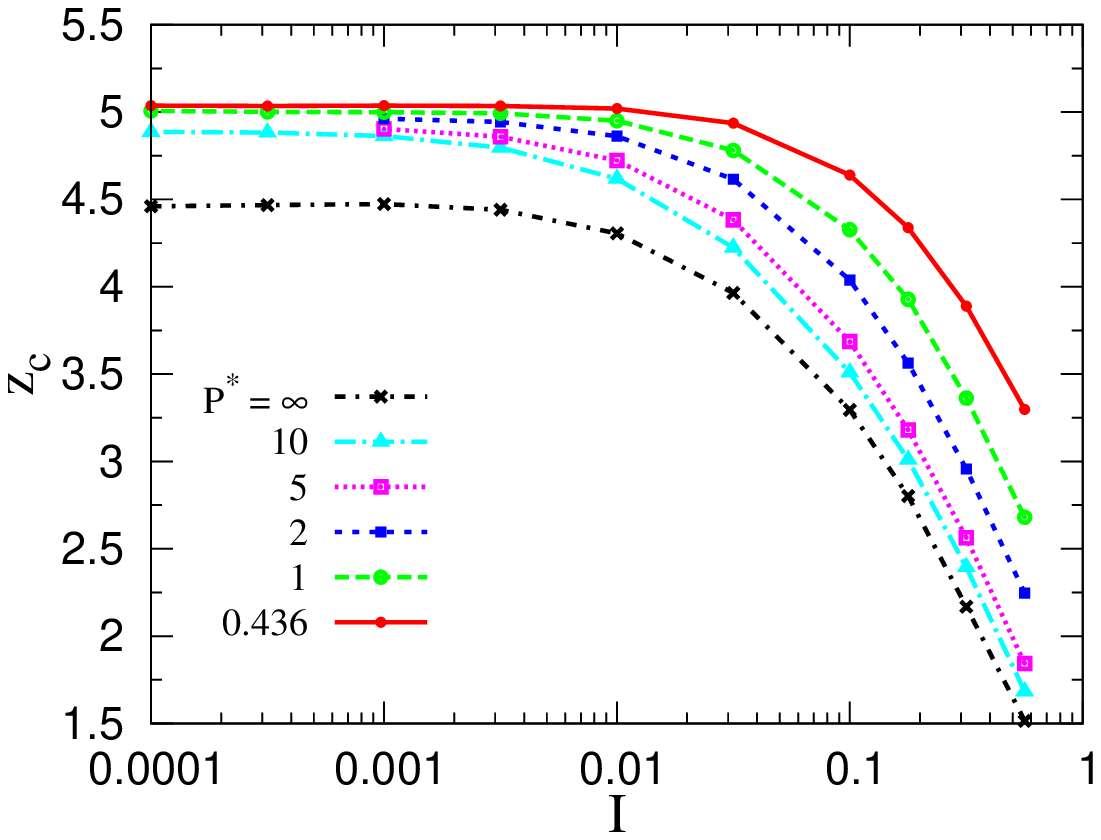} \\
  (b) \includegraphics[width=1.\linewidth]{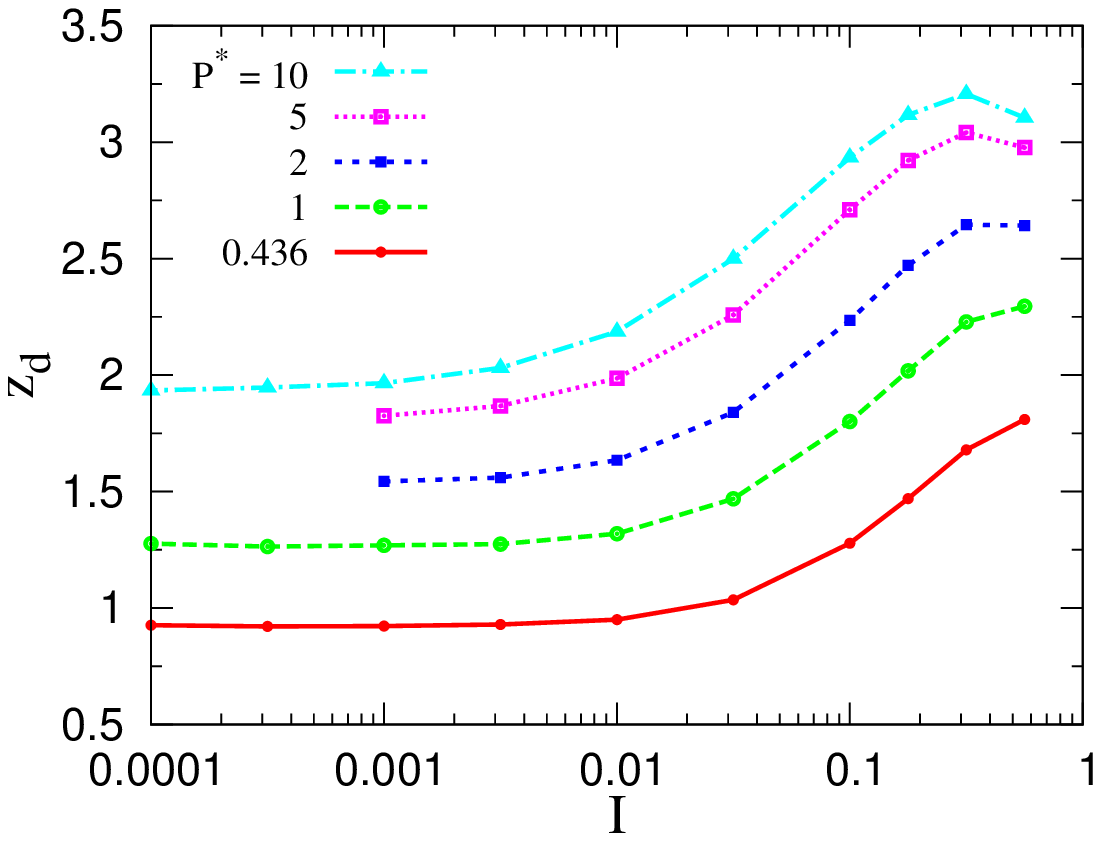}
  \caption{(Color online) Coordination numbers: (a) of contacts, $z_c$; (b) of distant interactions, $z_d$.}
\label{fig:cnC_cnD_I}
\end{figure}

\begin{figure}[htb]
  \centering
  \includegraphics[width=1.\linewidth]{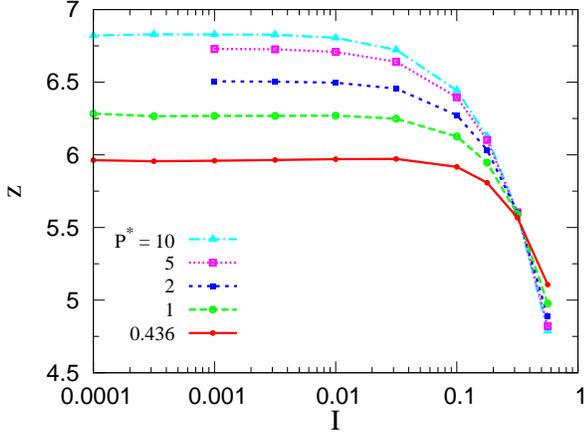} 
  \caption{(Color online) Total coordination number $z=z_c+z_d$.}
\label{fig:cn_I}
\end{figure}

Fig.~\ref{fig:cnC_cnD_I} shows the variations, with inertial number $I$, for different $P^*$ values, of coordination numbers $\zc$, 
for pairs of grains in contact, and $\zd$, for pairs of grains attracting each other without contact 
at a distance lower than $D_0$. The average number 
of contacts per grain, $z_c$, decreases for larger inertial numbers, as previously observed in 
cohesionless systems~\cite{Dacruz05,PR08a} and in cohesive ones~\cite{RRNC08}, slower for smaller $P^*$, as in~\cite{RRNC08} too. 
$z_c$ also increases as $P^*$ decreases at constant $I$, 
as previously observed as well~\cite{RRNC08}. Note that this latter trend is opposite to that of the 
solid fraction (Fig.~\ref{fig:mu_phi}). As the importance of adhesion, relative to confinement 
stresses, increases, looser systems are obtained, yet better coordinated. Grains tend to stick to 
one another, and may form loose aggregates, as in static or quasistatically compressed assemblies, 
for which little correlation is also observed~\cite{GiRoCa07,GiRoCa08} between density and 
coordination number. On the other hand, the variations of the coordination number of distant interactions, $z_d$, with both parameters $I$ and $P^*$,
are in the opposite direction to those of $z_c$ . As $I$ increases, 
so does $z_d$: contacting pairs tend to separate, but some remain bonded by liquid bridges. And for 
stronger cohesion (smaller $P^*$), $z_d$ is correlated with the system density. The faster approach 
to quasistatic limit at smaller values of $P^*$ is apparent in both figures. 
The fraction of \textit{rattlers} (beads carrying no force~\cite{PR08a}) in non-cohesive systems 
 is about $5\%$. In the cohesive case, due to the attractive forces, nearly all of the particles 
are bonded to others and the number of rattlers tends to zero, as observed 
in 2D simulation of cohesive powders~\cite{GiRoCa07,GiRoCa08}.
$z_d$ tends to compensate the changes of $z_c$, so that the total coordination number
$z=z_c+z_d$, throughout the investigated range of $I$ and $P^*$ values, exhibits rather small variations (see Fig.~\ref{fig:cn_I}).
Within the investigated parameter range, the maximum change in $z$, between 6.8 and 4.8, corresponds to a correction of internal friction  $\mu^*$, 
should we change the meniscus volume to maintain the total liquid volume constant, of about 5\% (see Tab.~\ref{tab:sat_eff}).
The contact coordination number does not change much with the force range or the meniscus volume. 
Setting $D_0=0$ (instead of the standard value 0.1 used in the present study) or decreasing 
the volume of the meniscus from its standard value $V=10^{-3}a^3$ down to $10^{-6}a^3$, merely leads to a small decrease of 
$\zc$, from 5 to 4.7 in the quasistatic limit, when $P^*=0.436$. However, it has a strong influence on $\zd$. 
Compared to the standard case, for $P^*=0.436$ and small values of $I$, it decreases from 0.9 down to a 
value below 0.3 when we set $D_0=0.01$, or down to about zero when we set $V=10^{-6}$. 

\begin{figure}
  \centering
  \includegraphics[width=1.1\linewidth]{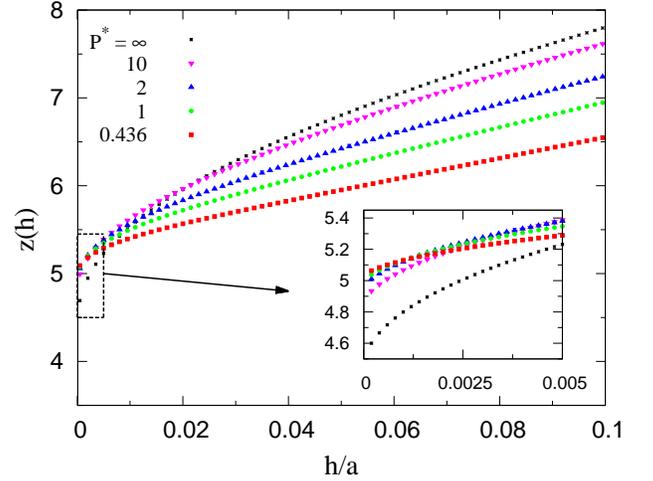}
  \caption{(Color online) Coordination number of neighbor grains versus  interparticle distance $h$, 
	   for different $P^*$ values and for $I=10^{-3}$. Inset shows detail at small $h$. } 
\label{fig:cn_h}
\end{figure}

It is interesting to compare the number of distant, interacting pairs to the total number of neighbor 
pairs at distance below $D_0$. The coordination number, $z(h)$, of neighbor grains at distance below 
$h$ (such that $z(0) = z_c$) grows with $h$ as depicted in Fig.~\ref{fig:cn_h}, 
corresponding to $I=10^{-3}$ (quasistatic limit).
$z(h)$, like the contact coordination number, is a decreasing function of $P^*$ for small $h/a$ 
(below about $2.5\times 10^{-3}$, see the insert in Fig.~\ref{fig:cn_h}). It increases with $P^*$, 
like the density, beyond that distance. In denser systems grains have more neighbors on average, but 
this is only true if neighbors at some distance are included in the count, and does not apply to 
contacts (a situation reminiscent of some observations in static packings of cohesionless 
grains~\cite{iviso1}). Up to meniscus rupture distance $D_0$, equal to $0.1a$ in the present case, 
each grain has on average $z(D_0)-\zc$ non-contacting neighbors, among which $\zd$ are joined by a liquid bridge.
Values of ratio $\zd/(z(D_0)-\zc)$ for different  
$P^*$ and $I$ are given in Tab.~\ref{tab:frac_lb}. The proportion of the 
neighbors within range $D_0$ that are bonded by a liquid bridge varies from 0.61 to 0.71 for 
$I=10^{-3}$ and between 0.68 and 0.79 for $I=10^{-1}$ -- slightly larger than the proportion 
$\sim 50$\% reported by Kohonen~\emph{et al.}~\cite{KOH04} in static grain packs.
\begin{table}[!htb]
    \centering
    \begin{tabular*}{.9\linewidth}{@{\extracolsep{\fill} } l c c|c c }
      \hline\hline
      \multicolumn{3}{c|}{$I=10^{-3}$} & \multicolumn{2}{c}{$I=10^{-1}$}\\
      \hline
      $P^*$ 	& $\zd$		& $\zd/(z(D_0)-\zc)$	& $\zd$		& $\zd/(z(D_0)-\zc)$ \\
      \hline
      0.436 	& 0.923		& 0.609		& 1.28		& 0.681 \\
      1		& 1.27		& 0.650 		& 1.80		& 0.723 \\
      2		& 1.54		& 0.675 		& 2.24		& 0.751 \\
      5		& 1.83		& 0.701 		& 2.71	        & 0.776 \\
      10	& 1.97		& 0.712 		& 2.93		& 0.787 \\
      \hline
    \end{tabular*}
    \normalsize
    \caption{Distant coordination number $\zd$ and proportion of pairs within distance $D_0$ joined by a meniscus,
	      $\zd/(z(D_0)-\zc)$, versus $P^*$ for two different values of $I$. \label{tab:frac_lb}}
\end{table}

If the meniscus forms as soon as grains approach at distance  $D_0$, rather than at contact, the number of contacts hardly changes ($\zc$  increases by 
about 5\% for $P^*=0.436$ in the quasistatic limit), but the increase in the number of menisci is larger than expected from the data of Tab.~\ref{tab:frac_lb}, from a simple
count of pairs within range $D_0$: $z_d$ is multiplied by  1.7 at small $P^*$ and $I$. 

\subsection{Pressure and average normal forces\label{sec:forcemoy}}
From Eq.~\ref{eq:stress}, neglecting the deflection of contacts in comparison to grain diameter $a$, and ignoring the kinetic term, 
one may relate~\cite{iviso1} the average pressure, $\cP=\tr \ww{\sigma}/3$, to the average normal force $\langle \Fn\rangle$ for all interactions, and to the average, $\langle \Fn h\rangle _d $, over pairs in distant interactions, of the product of force by distance $h\le D_0$:
\be
  \cP = {\Phi z \over \pi a^2} \langle \Fn\rangle + {\Phi \zd \over \pi a^3} \langle \Fn h \rangle _d
\label{eq:PvsFN}
\ee
Due to normal stress differences, ratio $\dfrac{\cP}{\sigma}_{\oyy}$ is only slightly different from  1 (about 0.95) at small $I$. 
We checked that formula~\eqref{eq:PvsFN}  is very accurate for all $P^*$ values, and found  its second term to be negligible, contributing less than 2\% of the pressure. 
\subsection{Contributions to stresses\label{sec:split}}
The contribution of the kinetic term in Eq.~\ref{eq:stress} to stresses is quite small. 
Even for the fastest flow in our simulation ($I=0.562$), this contribution does not 
exceed 2\% of the shear stress or 5\% of the normal stress components, and for $I=0.178$ 
it is nearly zero for all stress components. Therefore, in this section we only discuss 
the contributions of forces to the stress components, for the different values of the control 
parameters, $P^*$ and $I$. These contributions may be split in different ways, on 
distinguishing different forces. 
\subsubsection{Contact forces and distant capillary attraction}
\begin{figure}[h]
  \centering
    (a) \includegraphics[width=1.\linewidth]{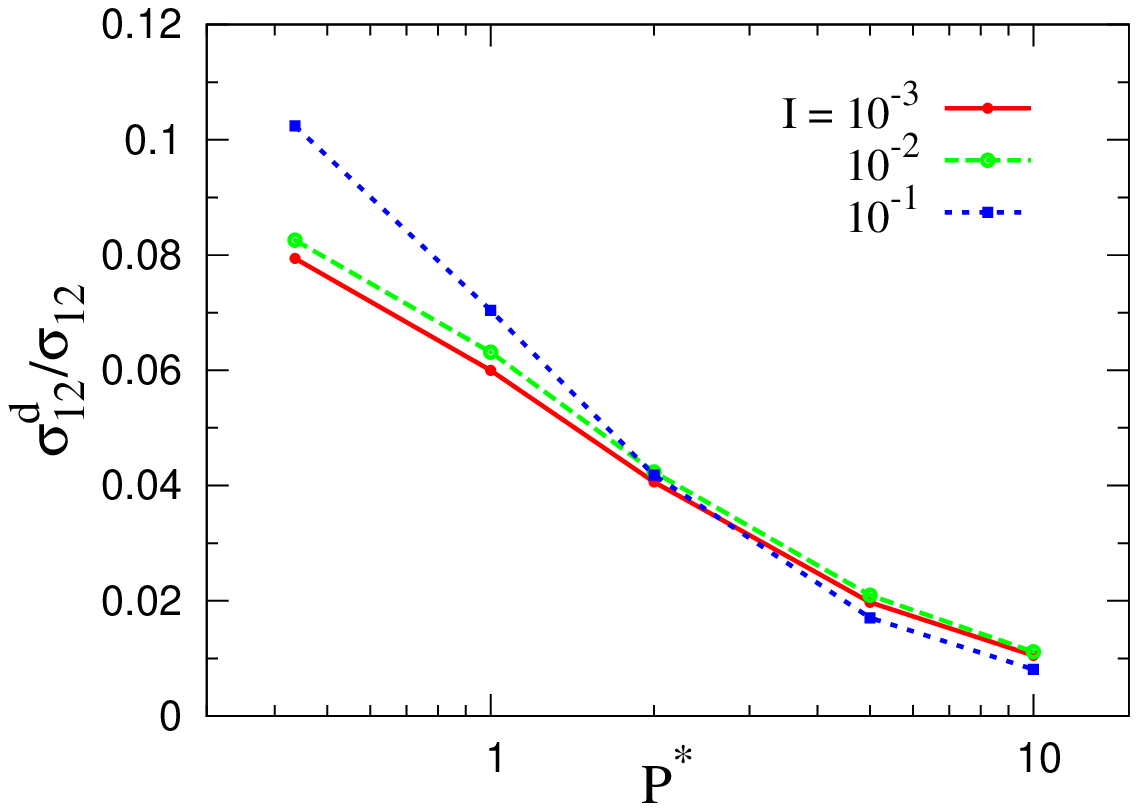}
    (b) \includegraphics[width=1.\linewidth]{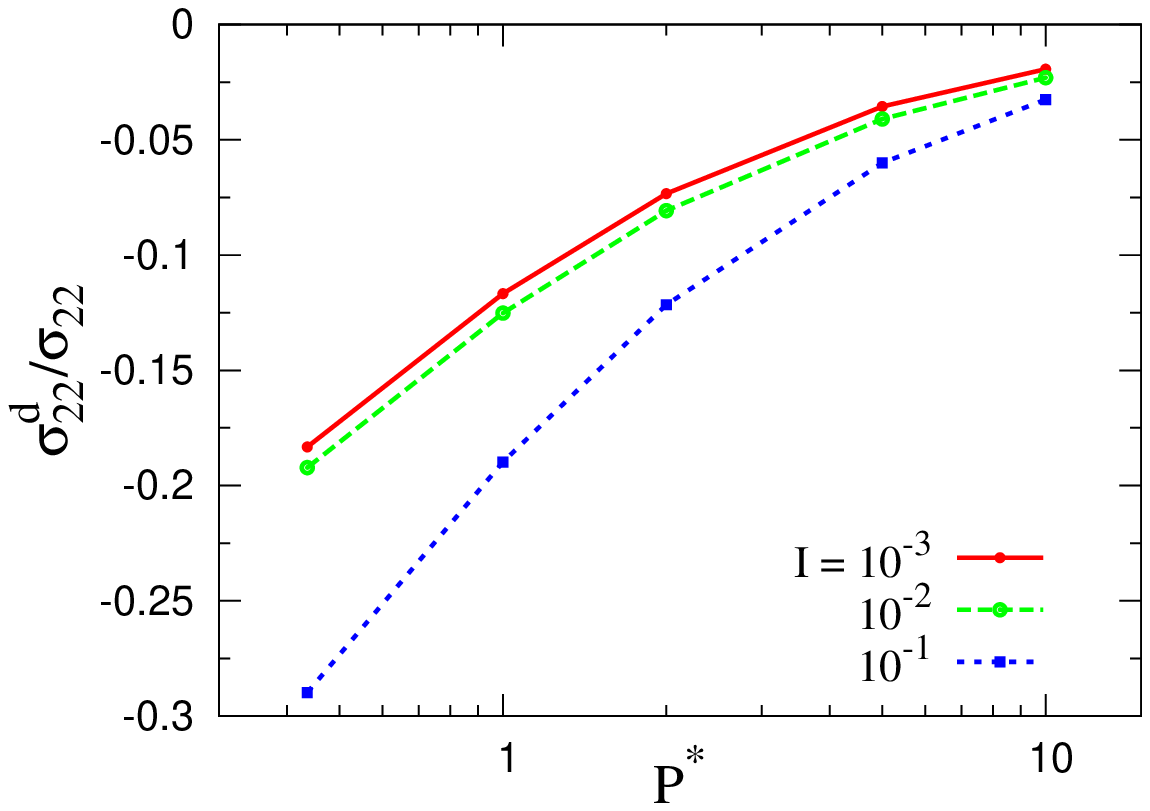}
  \caption{(Color online) Contribution of distant interactions to shear stress $\sigma_{\oxy}$ (a) and to normal stress $\sigma_{\oyy}$ (b).}
\label{fig:sxyd_syyd}
\end{figure}

First one may consider the total stress as a sum of the contributions of the contacts and of the distant interacting pairs, as
\be
  \sigma_{\alpha\beta}=\scon_{\alpha\beta}+\sd_{\alpha\beta}.
\label{eq:sxy_c_d}
\ee
Our results show that the contribution of contact forces  dominate in the shear stress. It is
larger than $90\%$, regardless of the values of $P^*$ and $I$. The contribution of distant 
interactions to $\sigma_{\oxy}$, as represented in Fig.~\ref{fig:sxyd_syyd}	a, is not negligible and 
increases for smaller values of $P^*$. However, it hardly exceeds $10\%$ of the total shear stress.

The contribution of distant interactions to $\sigma_{\oyy}$ is displayed in Fig.~\ref{fig:sxyd_syyd}b. 
Capillary forces being attractive, $\sd_{\oyy}$ is a tensile stress. 
For $P^*=0.436$, in the quasistatic limit, this contribution increases up to $20\%$ in magnitude. 
Consequently, the positive contribution of contact forces to $\sigma_{\oyy}$ reaches about $1.2\sigma_{\oyy}$ for 
$P^*=0.436$. 

The relative importance of the contributions of contacts and distant capillary forces to 
$\sd_{\oxx}$ and $\sd_{\ozz}$ is similar: in the quasistatic limit and for $P^*=0.436$, 
one has $\sd_{\oxx}/\sigma_{\oxx}\simeq -0.16$ and $\sd_{\ozz}/\sigma_{\ozz}\simeq -0.25$.

Accordingly, the contribution of normal contact forces is the dominant one 
in normal stress differences $N_1$, $N_2$ (with a notable contribution of tangential forces to $N_2$, typically 20\% at low $P^*$).
\subsubsection{Elastic-frictional forces and capillary forces}
An alternative decomposition of the stress tensor is:
\be
  \sigma_{\alpha\beta}=\scap_{\alpha\beta}+\sne_{\alpha\beta}+\st_{\alpha\beta},
\label{eq:sxyDecomp2}
\ee
in which $\scap_{\alpha\beta}$ is the contribution of capillary forces (either in the contacts or for distant interacting pairs), 
$\sne_{\alpha\beta}$ 
is the contribution of normal elastic forces and $\st_{\alpha\beta}$ is the contribution of tangential forces. 

The normal elastic forces contribute more than 90\% of the shear stress, whatever $P^*$ and $I$. 

\begin{figure}[h]
 \includegraphics[width=1.\linewidth]{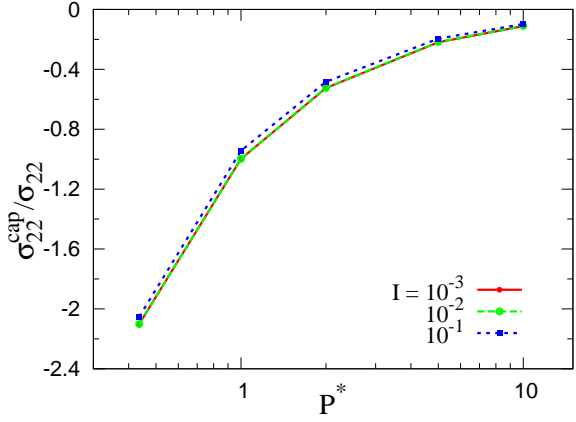}
  \caption{(Color online) Contribution of capillary forces to  stress $\sigma_{\oyy}$.}
\label{fig:syycapVsp}
\end{figure}

The contribution of tangential forces to the normal (diagonal) elements of the stress tensor is negligible, but that of
capillary forces is very important: for $P^*=0.436$,  negative terms
$\scap_{\alpha\alpha}$ ($1\leq\alpha\leq 3$) 
are very large in magnitude: one observes $\scap_{\alpha\alpha}<-2\sigma_{\alpha\alpha}$ for small $P^*$, as shown in Fig.~\ref{fig:syycapVsp}
This large negative contribution  is compensated by that of the repulsive normal elastic forces, $\sne_{\alpha\alpha}>3\sigma_{\alpha\alpha}$. 
Such a large negative contribution of capillary forces to pressure implies that the particles are strongly pushed against one another, which 
increases the sliding threshold for tangential contact forces. 

\begin{figure}[h]
  (a) \includegraphics[width=1.\linewidth]{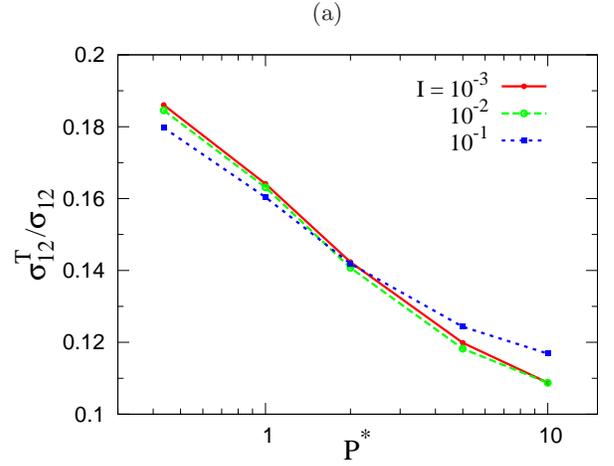}
  (b) \includegraphics[width=1.\linewidth]{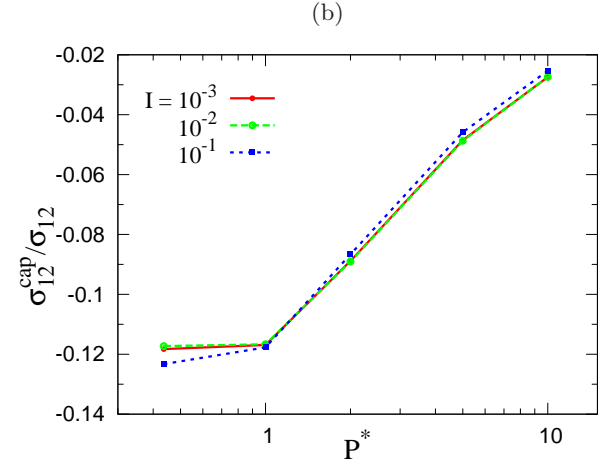}
  \caption{(Color online) Contributions of tangential (a) and capillary (b) interactions to total shear stress $\sigma_{\oxy}$.}
\label{fig:sxyMg_sxyT}
\end{figure}

Fig.~\ref{fig:sxyMg_sxyT} shows the contribution of tangential forces to the total shear stress. As $P^*$ is decreased to
$P^*=0.436$, the ratio $\st_{\oxy}/\sigma_{\oxy}$ increases to $0.18$. 
Capillary forces contribute to the shear stress with the opposite sign ($\scap_{\oxy}$ is positive while $\sigma_{\oxy}$ is negative). Fig.~\ref{fig:sxyMg_sxyT} 
shows that ratio  $\scap_{\oxy}/\sigma_{\oxy}$  is always negative and decreases down to -0.12 for $P^*=0.436$. 
Similarly to the case of normal stresses, the largest contribution is that of elastic normal forces, the (negative) contribution of which to $\sigma_{\oxy}$ compensates
the (positive) term $\scap_{\oxy}$ .
\subsection{Discussion\label{sec:discu}}
One important clue to understand the enhanced shear strength of the cohesive material, as compared to the cohesionless, dry granular assembly, is the large tensile contribution
of capillary force to normal stress:
\be
\scap_{\oyy} = -\beta\sigma_{\oyy},
\label{eq:defbeta}
\ee
with a coefficient $\beta$ ranging, in the quasistatic limit, 
 from about 0.15 ($P^*=10$) to 2.1 ($P^*=0.436$). Upon including the result for $P^*=0.1$ and $I=0.01$, $\beta$ reaches 
 about $7.2$. This coefficient, and its variations with $P^*$, can be approximately predicted from
the values of solid fraction and coordination numbers. 
Contacts ($z_c$, on average, per grain) carry capillary force $-F_0$, and distant forces ($z_d$ per grain)  average to a fraction of $-F_0$. 
Relation \eqref{eq:PvsFN} can be used to evaluate the capillary contribution to pressure $\cP$, as $-\dfrac{\Phi zF_0}{\pi a^2}\le \cP^{\text{cap}} \le -\dfrac{\Phi z_c F_0}{\pi a^2}$.
(This relation between  $ \cP^{\text{cap}}$ and contact tensile strength $F_0$ is sometimes referred to as the Rumpf formula, especially in the context of a prediction of 
rupture conditions~\cite{PAC98,RYR06,GiRoCa07}).  
Dividing by $\sigma_{\oyy}$, one obtains:
\be
-\frac{\Phi z}{\pi P^*}\le \frac{\cP^{\text{cap}}}{\sigma_{\oyy}} \le -\frac{\Phi z_c }{\pi P^*}.
\label{eq:pcap}
\ee
Ignoring the small difference between $\cP$ and $\sigma_{\oyy}$, \eqref{eq:pcap} provides an estimate of coefficient $\beta$ defined in~\eqref{eq:defbeta}. Thus 
the value of $\beta$ for reduced pressure $P^*=0.436$ is predicted between 1.9 and 2.3 (and for $P^*=0.1$, it should reach about 8).
Thus, quite unsurprisingly, the (negative) relative contribution of capillary forces  to normal stress if of order $ (1/P^*)\propto F_0/P$, with a coefficient 
that may be deduced from $\Phi$ and coordination numbers, according to \eqref{eq:pcap}.

\begin{figure}[h]
  \includegraphics[angle=270,width=.95\linewidth]{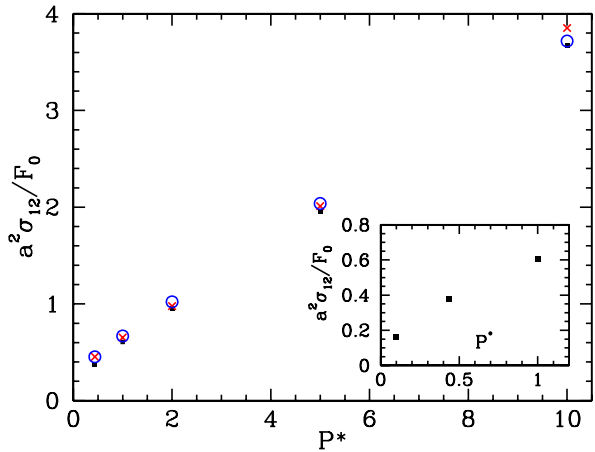}
  \caption{(Color online) $\sigma_{\oxy}$ versus  $\sigma_{\oyy}$, in quasistatic flow, in units of $F_0/a^2$. Square dots: numerical results (error bars are smaller); red crosses: predictions 
    of~\eqref{eq:frottpred1}, with exact coefficient $\beta$; blue circles: same with estimated $\beta$. 
    Insert: detail of numerical data for small $P^*$, including additional point at $P^*=0.1$. }
\label{fig:frott1}
\end{figure}

It is tempting to invoke a classical concept in geomechanics, that of \emph{effective pressure}, to describe the effect of capillary forces on the shear resistance of the material: 
the attractive forces create larger repulsive elastic reactions in the contact, corresponding to an effective pressure equal to $(1+\beta)\cP$. Furthermore, 
the local Coulomb condition in the contacts is to be written with those enhanced normal repulsive forces. Capillary forces also contribute to shear stress, but, as apparent in
Fig.~\ref{fig:sxyMg_sxyT}, in comparison to their influence on normal stresses, this is a small effect, and one may ignore it in a first approach.
One assumes then that the shear  behavior of the material is identical to that 
of a dry material under such effective normal stress $\sigma_{\oyy}^{\text{eff}}$. 
This approach leads to the following prediction for the $P^*$-dependent quasistatic friction coefficient $\mu^*_0 $:
\be
\mu^*_0 = (1+\beta) \mu_0^{\infty},
\label{eq:frottpred1}
\ee
in which $\mu_0^{\infty}$ denotes the quasistatic internal friction coefficient for dry grains, $P^*=\infty$.
Remarkably, if we further assume, as suggested by \eqref{eq:pcap}, that $\beta$ is roughly proportional to $1/P^*$, $\beta \simeq b/P^*$, we obtain a Mohr-Coulomb  
relation, Eq.~\ref{eq:mohrcoulomb}, for the stresses in the critical state:
with the same value of internal friction as in the dry case, $\mu^*_1=\mu_0^{\infty}$ and  a macroscopic cohesion given by
\be
c = \frac{b\mu_0^{\infty}F_0}{a^2}.
\label{eq:coh}
\ee
Fig.~\ref{fig:frott1} is a plot of 
$\sigma_{\oxy}$ versus  $\sigma_{\oyy}$ -- the yield locus -- in which the predictions of relation~\ref{eq:frottpred1}, 
both with the measured coefficient $\beta$ (Fig.~\ref{fig:syycapVsp}), 
and with the one predicted as $(z+z_c)\Phi/(2\pi P^*)$ from \eqref{eq:pcap},  are confronted with the numerical results. 
The admittedly crude prediction of relation~\eqref{eq:frottpred1} appears surprisingly close to the numerical results on this plot. The relative error in the prediction for stress ratio
$\mu_0^*$, with the measured value of $\beta$, is actually about 5\% at $P^*=10$, increasing to 20\% at $P^*=0.436$,  
and the value of $\mu_0^*$ for $P^*=0.1$ ( $\simeq 1.6$) from the measurements at $I=0.01$ is largely overestimated, at 2.7. 

One may  directly test for the validity of a Mohr-Coulomb relation to the data by fitting a linear form for the data of Fig.\ref{fig:frott1}.
Given the error bars (which are small and do not appear on the graph), 
an attempted  straight line fit through all 5 data points with $P^*\ge 0.43$  in Fig.~\ref{fig:frott1} is unambiguously rejected by the standard likelihood criterion. A linear fit
 is (barely) acceptable upon ignoring the value $P^*=0.436$, yielding  $\mu^*_1=0.340 \pm 0.001$ and $a^2c/F_0 =0.267\pm 0.005$ for the Mohr-Coulomb parameters.
From  \eqref{eq:frottpred1}  the predicted apparent macroscopic cohesion is above $0.3F_0/a^2$, and varies according to which data are used to identify $b$ in~\eqref{eq:coh}.
The result $\mu^*\simeq 1.6$ for $P^*=0.1$ (corresponding to $a^2\sigma_{\oxy}/F_0=\mu^* P^*=0.16$) is thus in contradiction with the 
Mohr-Coulomb model, which becomes increasingly inadequate for smaller $P^*$, as  apparent in the insert in Fig.~\ref{fig:frott1}.
 
\begin{figure}[h]
  \includegraphics[angle=270,width=.9\linewidth]{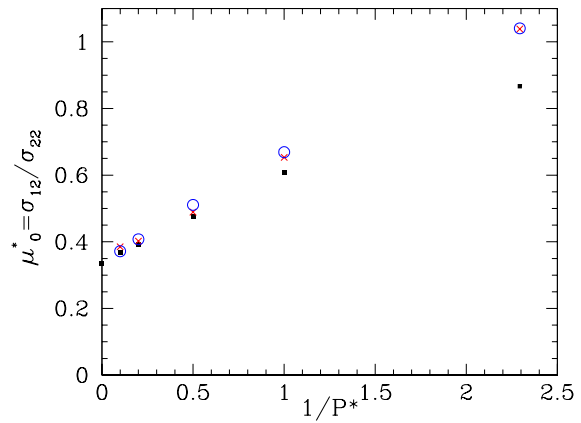}
  \caption{(Color online) Apparent quasistatic friction coefficient $\mu^*_0$ versus $1/P^*$ -- showing the value of $\mu^{\infty}_0$ for $1/P^*=0$.
   Measurements and predictions of~\eqref{eq:frottpred1}, with exact and estimated coefficient $\beta$, same symbols as in Fig.~\ref{fig:frott1}. }
\label{fig:frott2}
\end{figure}

The performance of the simple effective pressure prediction for the  $P^*$ dependence of $\mu^*_0$
 is better visualized in Fig.~\ref{fig:frott2}, which, unlike Fig.~\ref{fig:frott1}, is not sensitive to stress scale. 
 The global increase of $\mu^*_0$ is predicted, yet overestimated for the smallest $P^*$ values.
 
 One aspect that is not captured by this approach is the dependence of $\mu^*_0$ on meniscus volume (Tab.~\ref{tab:sat_eff}): 
 the variations of coefficient $\beta$ (from 1.8 to about 2.2
 as $V$ increases from $10^{-6}a^3$ to $5.10^{-3}a^3$) are  insufficient to account for the increase of the friction coefficient.  
 
There are quite a few reasons for the effective pressure approach to fail: while the  mechanical properties are supposed to be the same once stresses are corrected, 
the density of the material, for one thing, is 
different in the dry and the wet case (with $\Phi$ varying between $0.525$ and $0.595$ as $P^*$ grows from 0.46 to infinity); capillary forces also contribute to shear stress,
the force network is bound to be different, etc.
Nevertheless, although admittedly crude, 
 the prediction based on \eqref{eq:frottpred1} proves apt to capture the trend of the change of $\mu^*_0$
 with $P^*$, although it overestimates its growth at small $P^*$. As to the Mohr-Coulomb representation of yield stresses, it might be used as an  approximation for
 $P^*\ge 1$, but the observations clearly preclude the definition of unique values of macroscopic cohesion and  friction coefficient according to~\eqref{eq:mohrcoulomb} for smaller
 pressures.
 
 Ref.~\cite{PAC98} reports on a  laboratory study of quasistatic yield loci ($\sigma_{\oxy}$ versus $\sigma_{\oyy}$ at the onset of plastic yielding and flow) of various kinds of 
 wet granular assemblies in the pendular regime, including glass beads, which 
 offer a suitable experimental comparison to our results. It proposes (under the name 'shift theory'), exactly the same effective pressure approach as the one we have
 attempted here, and concludes that it provides a good approximation, by which the yield condition of wet materials is deduced from the one of the dry grains. Interestingly, the 
 investigated $P^*$ values in this study range from about 0.2 to $\sim 2.5$, and the yield locus is slightly concave, as in our numerical results. Measured values of $\mu^*$
 are similar to our results (with e.g., $\mu^* \simeq 0.7$ for $P^*=1$), and little change is obtained by increasing saturation by a factor of 3. 
 Some possible differences between those experiments and our simulations could result from the different state of the material (the
  experiments are not necessarily carried out in steady state quasistatic shear flow, and could depend on the initial assembling process), and from the value of the wetting angle.
  However, quite a satisfactory semi-quantitative agreement is obtained. 

In the following sections, for a better assessment of the rheophysical effect of attractive capillary forces, 
microscopic and microstructural aspects of force networks are investigated in greater detail.
\section{Force distribution\label{sec:fn_unsat}}
\begin{figure}[h!]
  \centering
 \includegraphics[width=1.05\linewidth]{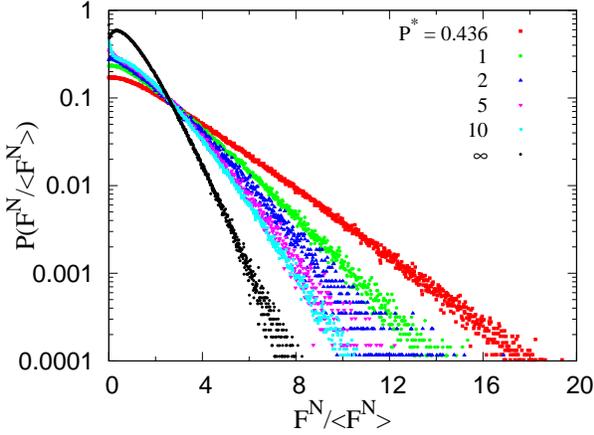}
  \caption{(Color online) Distributions of normal forces normalized by average normal force $\langle\Fn\rangle$ for $I=10^{-3}$ and different values of $P^*$.}
\label{fig:dist_FNvsP_ND}
\end{figure}

The distribution of  intergranular force values in a granular material in equilibrium~\cite{RJMR96,SGL02,iviso1,DTS05,PR08b}, or in inertial flow~\cite{Dacruz05,PR08a} has received a lot of attention in the recent literature. While the probability distribution function of force values in cohesionless systems tends to decrease exponentially, on a scale given by the average $\langle F_N\rangle$, in cohesive granular assemblies, characterized by the contact tensile strength $F_0$, the equilibrium force distribution evolves, as $P^*$ decreases to low values, towards a 
roughly symmetric distribution about zero, with values of both signs of order $F_0\gg \langle F_N\rangle$~\cite{GiRoCa07,GiRoCa08}. As compared to the two-dimensional 
results of Refs.~\cite{GiRoCa07,GiRoCa08}, the present 3D numerical study of wet spherical grain assemblies does not investigate 
very small $P^*$ states, but involves longer-ranged distant interactions. The positive force wing of the p.d.f. of normal forces near the quasistatic limit 
is shown in Fig.~\ref{fig:dist_FNvsP_ND},
showing the gradual departure from the cohesionless distribution shape, and the transition to a cohesion-dominated force network with values of order $F_0$, ratio
$\langle F_N\rangle/F_0$ being approximately proportional to  $P^*$ as discussed in Sec.~\ref{sec:forcemoy}.

\begin{figure}[h!]
  \centering
  \includegraphics[width=1.05\linewidth]{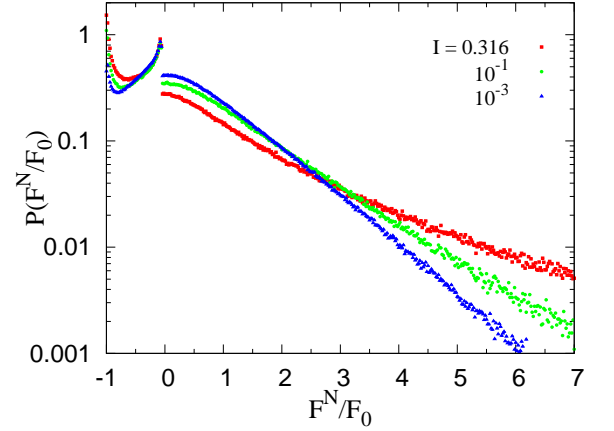} 
  \caption{(Color online) Distributions of normal forces for $P^*=0.436$ and different values of $I$, normalized with $F_0$.
\label{fig:DistFNF0}
}
\end{figure}

\begin{figure}[h!]
  \centering
   \includegraphics[width=1.05\linewidth]{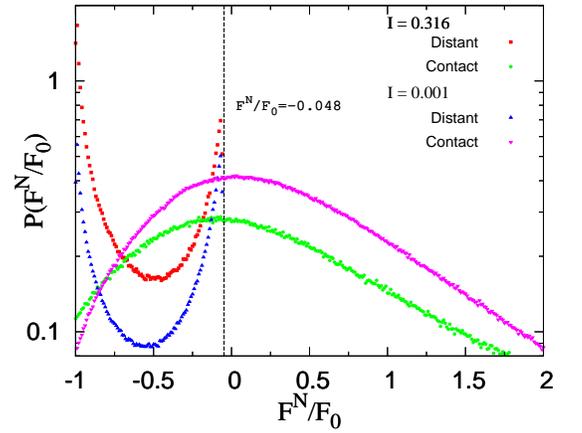}
  \caption{(Color online) Contributions of contact and distant interactions to p.d.f. of  normal forces  for $P^*=0.436$,  and 
  two values of $I$, $I=0.316$ and $I=0.001$. The vertical dashed line corresponds to force at rupture distance, $\fcap(D_0)$.}
\label{fig:dist_FNC_FND}
\end{figure}

At low reduced pressure, as for $P^*=0.436$, it is more appropriate to normalize the distribution by  $F_0$, as in Fig.~\ref{fig:DistFNF0}. This plot shows the influence of
inertia parameter $I$, which is, for large positive values, qualitatively similar to the one observed with dry grains: the distribution widens, large forces being associated with
collisions between grains or groups of grains.  
Another effect of increasing the inertial number is, as expected from the results of Fig.~\ref{fig:cnC_cnD_I}, a depletion of the population of contacts, compensated by
a greater number of distant grains joined by a meniscus. To understand better the distribution shape for negative values, Fig.~\ref{fig:dist_FNC_FND} distinguishes the distributions
of contact and distant (attractive) forces. Contact force distributions exhibit a maximum in zero, with negative values becoming more frequent as $I$ increases. The larger value of the pdf near $-F_0$ signals then the opening of more contacts. 
The distant interactions are responsible for the non-monotonic part of the pdf. On the one hand, 
the sharp maximum near $-F_0$ signals a large population of grain pairs at close distance, 
in agreement with the fast increase of $z(h)$ at small $h$ visible in Fig.~\ref{fig:cn_h}. 
On the other hand, the increase near the minimum attractive force at rupture distance $D_0$ 
merely reflects the slow variation of function $\fcap(h)$ (Fig.~\ref{fig:Force_model}).

The "effective pressure" concept relies on the assumption that the effect of attractive capillary forces are similar to that of a larger applied isotropic stress.
One way to test such an idea at the microscopic scale is to compare the distributions of normal elastic forces: if normalized by the average elastic force, related to the effective pressure, those should be independent on $P^*$ and similar to the force distribution in a cohesionless system.
\begin{figure}[h!]
  \centering
   \includegraphics[width=1.05\linewidth]{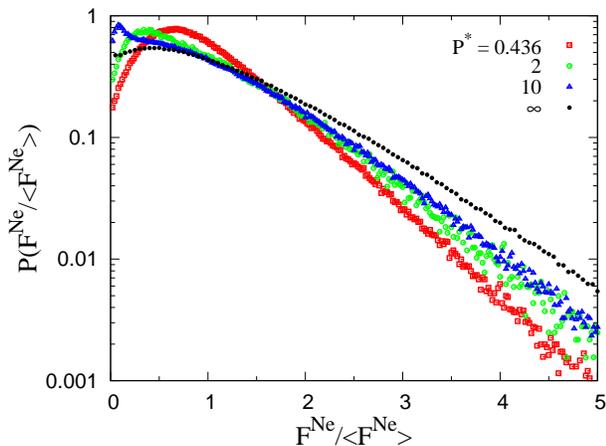}
  \caption{(Color online) Distributions of normal elastic forces at small $I$ for different $P^*$ values.}
\label{fig:dist_FNe}
\end{figure}
Fig.~\ref{fig:dist_FNe} compares the distributions of elastic normal forces, normalized by their mean value, for small $I$ and different values of $P^*$. 
Those distributions are  roughly similar, but show, as expected, notable discrepancies for values of order $F_0$. The decay for large values is faster in cohesive systems, reflecting 
a difference in force networks.
%
%
 \section{Agglomeration\label{sec:aggl}}
The aggregation of cohesive grains is observed and reported in many numerical and experimental studies, and is exploited in industrial processes~\cite{RRNC08,TTO01}.
It was directly observed  in flow of cohesive granular assemblies, both in numerical model materials~\cite{RRNC08}, and in experiments with wet powders~\cite{MSWK00}.
A numerical study of steady state chute flow ~\cite{BGLL05}
reports an increase of the number of long-lasting contacts in the presence of cohesive forces. 
Weber \textit{et al.} in~\cite{WHH04}, carried out a 
detailed study of the effect of capillary forces on agglomerate duration and size.
The agglomeration phenomena in steady shear flow 
 is studied here, first, by measuring contact ages and meniscus ages, depending on state parameters. 
Then, the age-dependent size of clusters is measured, depending on $P^*$ and $I$.
These clustering properties are related to the material rheology.
 \subsection{Age of contacts and of distant interactions}
\begin{figure}
  \centering
  \includegraphics[width=1.05\linewidth]{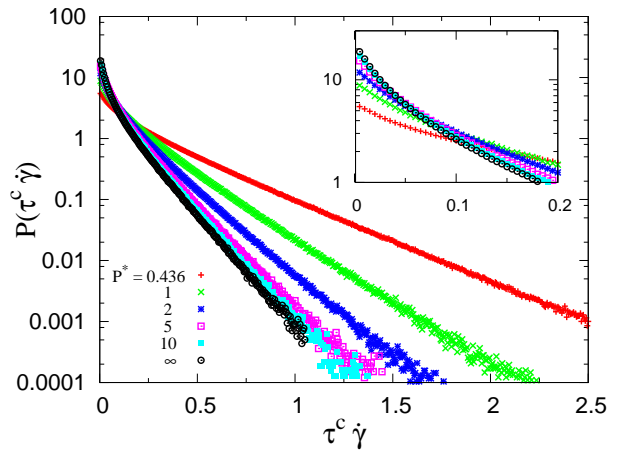}
  \caption{(Color online) Distribution of the age of contacts for different values of $P^*$ and $I=0.1$. Inset shows the same graph in a shorter range of $\tauc\dot\gamma$.}
\label{fig:ageC_P}
\end{figure}

\begin{figure}
  \centering
  \includegraphics[width=1.05\linewidth]{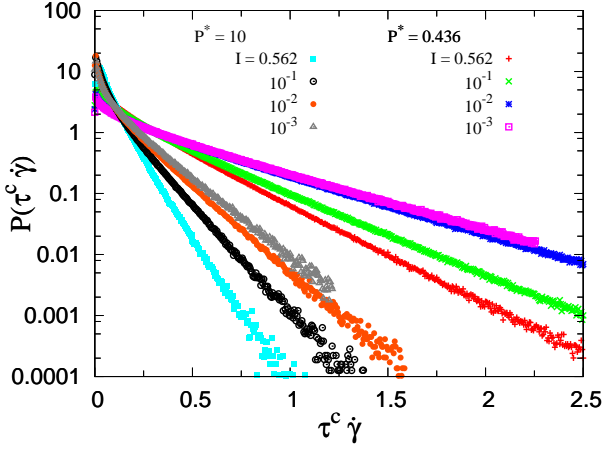}
  \caption{(Color online) Distribution of the age of contacts for different values of $I$ and two different values of $P^*$.}
\label{fig:ageC_I}
\end{figure}

\begin{figure}
  \centering
  \includegraphics[width=1.05\linewidth]{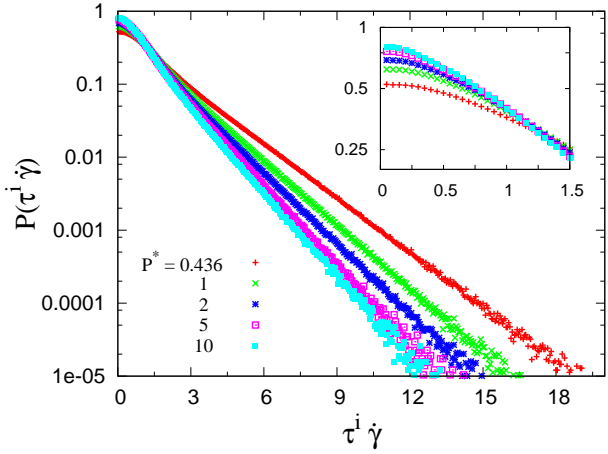}
  \caption{(Color online) Distribution of the age of menisci for different values of $P^*$ (same for all values of $I$). Inset shows the same 
  graph in a shorter range of $\taui\dot\gamma$.}
\label{fig:ageD_P}
\end{figure}

The distribution of the age $\tauc$ of contacts for $I=10^{-1}$
and different values of $P^*$ is shown in Fig.~\ref{fig:ageC_P}.
$P(\tauc\dot\gamma)$ is the probability distribution of contact ages $\tau_c$, expressed as a  strain~$\tauc\dot\gamma$. 
The decrease of $P(\tauc\dot\gamma)$ is slower for smaller $P^*$, showing that for the stronger cohesive forces 
the contacts survive over larger strain intervals~\cite{BGLL05,WHH04}. For large enough strains, 
$\tauc\dot\gamma>0.5$, these probability density functions decay with an exponential form, $P(\tauc\dot\gamma) \propto e^{-\tauc/\tauz}$. 
Values of decay times $ \tauz$, given in Tab.~\ref{tab:age_P}, 
increase as we decrease $P^*$. Average contact ages, $\tauc_{\text{avg}}$, also provided in the table, show the same behavior
($\tauc_{\text{avg}}$ is smaller than $\tauz^{\text{c}}$ because the distribution is not exponential for short times -- see inset on the figure).
Fig.~\ref{fig:ageC_I} shows the evolution of the pdf with $I$ for two different values of $P^*$, revealing, as expected, that 
contact ages (in units of $1/\dot\gamma$) decrease in faster flows. 
For $I\leq 10^{-2}$,  curves appear to coincide, showing nearly quasistatic behavior. 
The probability distribution function of the age of interactions $P(\taui\dot\gamma)$ (i.e., the age of liquid bridges) 
is also shown in Fig.~\ref{fig:ageD_P} for different values of $P^*$. 
Liquid bridges survive for quite large strain intervals, reaching several units with a probability of order 0.1, which increase as $P^*$ decreases. 
Initially, most liquid bridges survive at least for strains of order $0.1$. Beyond unit strain  curves might be fitted by an exponential function too, 
defining a decay time $\tauz^{\text{i}}$. $P^*$-dependent values  of 
$\tauz^{\text{i}}$ and of the average meniscus age  $\taui_{\text{avg}}$ are listed in Tab.~\ref{tab:age_P}. Remarkably,  the curves do not present any notable difference
for different values of $I$: the pairs may lose their contacts in faster flows, but they are still bonded with liquid bridges.
The age of contacts and of distant interactions thus reveal the formation 
of aggregates in the presence of capillary forces. These clusters 
are transported by the flow for some distance before they are broken or restructured.
They may survive for strain intervals of a few units.
\begin{table}[!htb]
    \centering
    \begin{tabular*}{0.6\linewidth}{@{\extracolsep{\fill} } lcccc}
      \hline\hline
      $P^*$	& $\dot\gamma\tauz^{\text{c}}$	& $\tauc\dot\gamma_{\text{avg}}$ & $\dot\gamma\tauz^{\text{i}}$ & $\taui\dot\gamma_{\text{avg}}$ \T\\
      \hline
      $0.436$	& 0.306			& 0.258					& 1.704		& 1.609 \T\\
      $1$	& 0.180 		& 0.154					& 1.437		& 1.325 \\
      $2$	& 0.153 		& 0.111					& 1.295		& 1.187 \\
      $5$	& 0.128 		& 0.087					& 1.164		& 1.072 \\
      $10$	& 0.120 		& 0.080					& 1.102		& 1.021 \\
      $\infty$	& 0.118 		& 0.074					& ---		& ---\\
     \hline\hline
    \end{tabular*}
    \normalsize
    \caption{(Color online) Decay time of age distribution function for contacts,  $\tauz^{\text{c}}$, and for
    all interactions, $\tauz^{\text{i}}$, obtained by an exponential fit to the data of
    Fig.~\ref{fig:ageC_P} and of Fig.~\ref{fig:ageD_P}); 
    average contact age $\tauc_{\text{avg}}$ and interaction age $\taui_{\text{avg}}$,
    for different values of $P^*$ and $I=0.1$. All four times are normalized by shearing time $1/\dot\gamma$.}
\label{tab:age_P}
\end{table}
\subsection{Clusters}
\begin{figure}[H]
  \centering
  \includegraphics[width=1.05\linewidth]{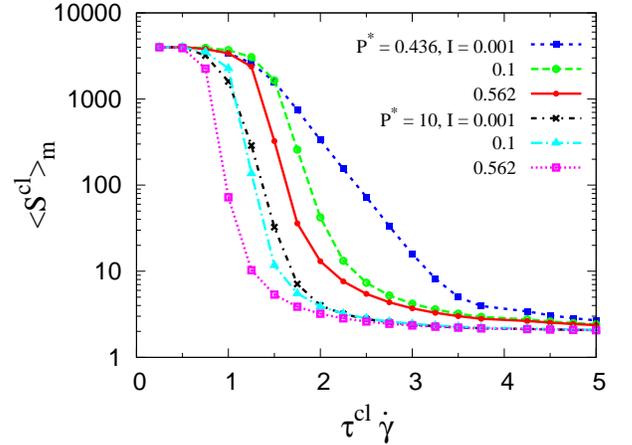}
  \caption{(Color online) Mass-averaged cluster size, $\ave{\scl}_{\text{m}}$, versus cluster age $\taucl\dot\gamma$ for different values of $P^*$ and $I$.}
\label{fig:mass_avg}
\end{figure}

Clusters are defined
as sets of grains connected by liquid bonds  for a minimum time, $\taucl$, and the clustering tendency might be appreciated on recording
the $\taucl$-dependent mass-averages, cluster size:
\be
  \ave{\scl}_{\text{m}}= \frac{\sum_i S_i^2}{\sum_i S_i},
\label{eq:massAvg}
\ee
in which the summations run over all clusters $i$ containing $S_i$ grains. 
The results for $\ave{\scl}_{\text{m}}$ for the different values of $P^*$ and $I$ are shown
in Fig.~\ref{fig:mass_avg}, as functions of $\taucl\dot\gamma$. 
For small values of $\taucl\dot\gamma$ almost all particles are gathered in a single cluster, which results in 
$\ave{\scl}_{\text{m}}\simeq 4000$. For large values of $\taucl\dot\gamma$ most
particles are isolated or part of a very small cluster, and so a value of order 1 for $\ave{\scl}_{\text{m}}$ is expected. 
In the midrange of 
$\taucl\dot\gamma$, $\ave{\scl}_{\text{m}}$ increase for decreasing  $P^*$ or $I$, i.e. 
for stronger capillary forces or slower flows.
\section{Fabric anisotropy\label{sec:aniso_unsat}}
The capacity of granular assemblies to form anisotropic force networks is the only origin of shear strength with frictionless grains~\cite{PR08a,PR08b,ARR15}, and is 
known to play a central role in the shear strength of frictional grains as well. 
To understand how contact and distant capillary forces contribute to the shear stress, it is instructive
to study the distribution of contact orientations (normal vectors $\bs n$ on the unit sphere $\Sigma$), $E(\bs n)$, which, to lowest order, is characterized by the fabric tensor: 
\be
F _{\alpha\beta} = \langle n_\alpha n_\beta\rangle = \int_\Sigma E(\bs{n}) n_\alpha n_\beta \,d^2 \bs n 
\label{eq:fabric}
\ee
\begin{figure}
  \centering
  \includegraphics[width=8.0cm]{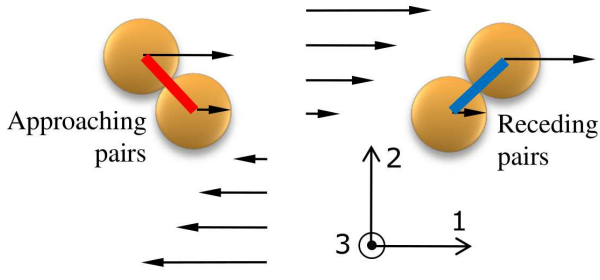}
  \caption{(Color online) Sketch of approaching ($\varphi$ near $3\pi/4$) and receding pairs ($\varphi$ near $\pi/4$) 
  in macroscopic shear flow.}
\label{fig:orientCD}
\end{figure}
The connection between fabric and normal force contribution to stresses, or $ \ww\sigma^{\text{N}} $, is quite direct, as one
has:
\be
  \sigma^{\text{N}}_{\alpha\beta}= \frac{z \Phi}{3\pi a^2} \int_{\Sigma} E(\bs{n}) \langle \Fn \rangle_{\bs n} \hspace{0.2em} n_{\alpha} n_{\beta} \,d^2 \bs n ,
\label{eq:stress2}
\ee
an integral over the unit sphere in which $\langle \Fn \rangle_{\bs n}$ denotes the average normal force carried by the pairs with orientation $\bs n$. 
As reported in Sec.~\ref{sec:split}, the contribution of the normal forces, 
$ \ww\sigma^{\text{N}} $, amounts to more than 80\% of the shear stress. 
The contribution of fabric parameters $F_{\oxy}$  to shear stress $\sigma_ {\oxy}$ might be visualized in
Fig.~\ref{fig:orientCD}. On average, if pairs are preferentially oriented with the normal vector 
 within a compression  quadrant in  the shear flow, then $F_{12}<0$ will tend to increase the absolute value of  $\sigma_{\oxy}$
 if forces are positive, and to decrease it they are negative. On the other hand, negative forces will increase the absolute value  of  $\sigma_{\oxy}$ if 
 preferentially oriented in the extension quadrants in the shear flow. 
 \begin{figure}[h]
  \centering
  (a)  \includegraphics[width=1.\linewidth]{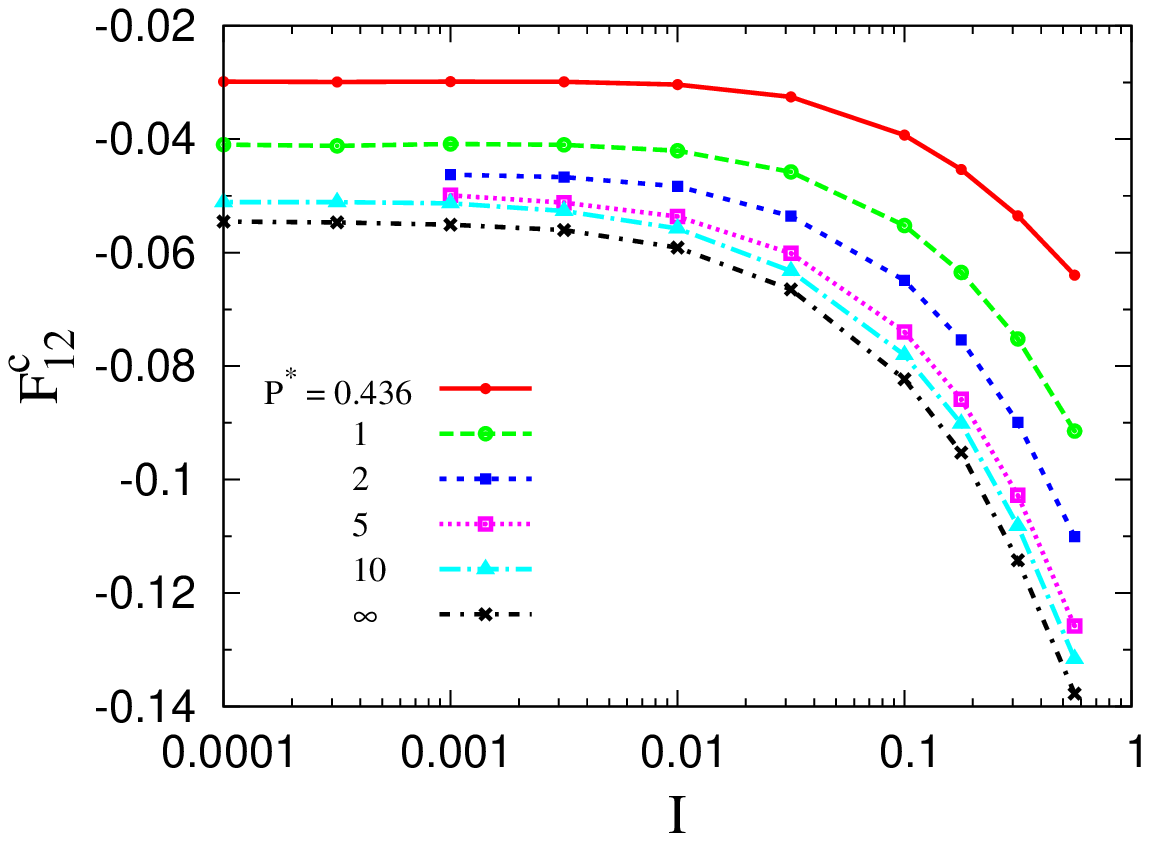} \\
  (b)  \includegraphics[width=1.\linewidth]{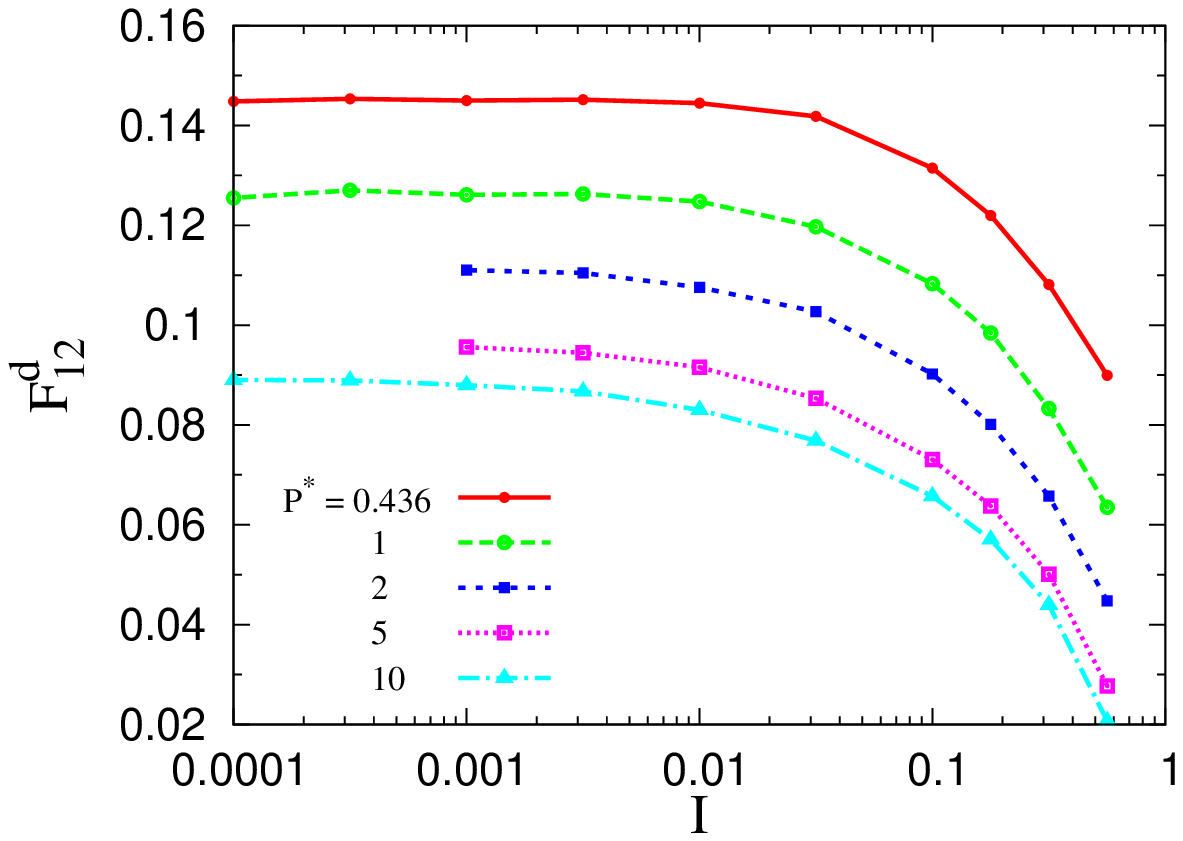}
  \caption{(Color online) Fabric parameter $F_{12}$ for contacting pairs (a) and distant interactions (b) 
    versus $I$ for different $P^*$.}
\label{fig:Fc12_Fd12_I}
\end{figure}

The evolutions of fabric parameters $\Fc_{\oxy}$ and $\Fd_{\oxy}$, pertaining respectively to contact and distant normal forces,  versus $I$, for different $P^*$ 
values, are displayed in Fig.~\ref{fig:Fc12_Fd12_I}. $\Fc_{\oxy}$ is negative, signaling the contribution
of normal contact forces to shear strength (as $\langle\Fnc \rangle$ is the dominant contribution to $\langle\Fn \rangle$, related to $\cP$ by \eqref{eq:PvsFN}). 
The largest value, and the largest variation of $\Fc_{\oxy}$ with $I$, is obtained in the dry system ($P^*=\infty$). The decrease of this anisotropy parameter  for smaller $P^*$ 
 may be understood in reference to the clustering phenomena and to the larger duration of contacts evidenced in Sec.~\ref{sec:aggl}. Longer-lived contacts rotate
in the shear flow, and are less favorably oriented in the compression quadrant. Shear flow carries  agglomerates for some distance before they break and thus their 
random tumbling motion increases the isotropy of the contact orientations. 
In faster flows (for larger $I$), while contacts tend to open in cohesionless materials,  
 enhancing fabric anisotropy, cohesive contacts can resist flow agitation and inertial effects better, whence a smaller $I$ influence; if they open, they transform into attractive distant interactions, and the anisotropy of distant interactions also decreases. Those distant capillary forces are characterized by a comparatively large anisotropy, about three times as large as $\vert\Fc_{\oxy}\vert$. As $\Fd_{\oxy}$ is positive,  those distant attractive forces contribute to increase the internal friction coefficient. Unlike  $\vert\Fc_{\oxy}\vert$, $\Fd_{\oxy}$ increases for smaller $P^*$ values, which corresponds to the growing contribution of distant interactions to shear strength shown in Fig.~\ref{fig:sxyd_syyd}. The
 different rule of meniscus formation (at contact) and breakage (at distance $D_0$) explains, in part, this large fabric anisotropy of attractive forces: approaching particles
 are not attracted to each other, whence a small number of distant interacting pairs in the compressive quadrant, with a negative contribution to $F_{12}$; as particles get
 separated, receding pairs are still attracted to each other, whence a positive contribution to $F_{12}$ from the extension quadrant. In the model without meniscus hysteresis,
 assuming capillary attraction to appear as soon as grains approach within distance $D_0$, $\Fd_{\oxy}$ strongly decreases, from 0.14 to about 0.07 at $P^*=0.436$ and small $I$.


\section{Summary and discussion\label{sec:conc}}
The rheological properties of unsaturated granular materials, 
in which a small amount of wetting liquid, forming  liquid bridges  and transmitting attractive capillary forces between particles,  
generalize, in many respects, 
previous observations on cohesive granular materials, with macroscopic properties exhibiting similar dependences on  $I$ and $P^*$.
 Thus, compared to dry materials,  the apparent internal friction coefficient $\mu^*=\sigma_{12}/\sigma_{22}$ is enhanced (from 0.33 to more than 1 in the explored range 
 $P^*\ge 0.1$);  looser structures  are stabilized, 
 even in the quasistatic limit ($\Phi \simeq 0.52$, for $P^*=0.436$ is below all packing densities with cohesionless grains), even though contact coordination numbers, due to the absence of
 rattlers, may be larger. Our results describe those effects in quantitative form, in the range $P^*\ge 0.1$ and $10^{-4}\le I\le 0.56$, and 
 specify the dependence on various features of the model. We only predict quite 
 a small dependence of the rheology on saturation within the pendular range (up to 5--10\%), in agreement with experimental observations~\cite{PAC98,RYR06}.
 
 More accurate models of the capillary force
 dependence on intergranular distance, or of the distribution of liquid between menisci with varying volumes, would hardly change the results. 
 Interestingly, though, some variants of the 
 model, although arguably not realistic,  have  notably different  rheological properties. 
 Thus, reducing the meniscus volume to very small values would have quite a notable effect on 
 internal friction and density -- but, in practice, menisci are  unlikely to form with such small liquid contents. Assuming menisci form as soon as grains approach to their maximum extension distance (range of capillary force) would also strongly affect macroscopic properties. 
 
Shear localization systematically affects shear flows at low $P^*$, and we could not measure the constitutive 
behavior at $P^*= 0.1$ except for some intermediate $I$ values of order 0.01. Localized states are characterized by  velocity profiles with gradients concentrated within narrow bands, where the solid fraction is  well below its bulk value. The band thickness lies in the range of 5 to 10 grain diameters $a$ at small $I$, but might be as small as about $1.5 a$ in faster
flows ($I\ge 0.1$). 

We also record normal stress differences, which are larger than for dry grains, and tend to grow with decreasing $P^*$. 
The second normal stress difference, in particular, reaches 20\% of the imposed normal stress $\sigma_{22}$ in the quasistatic limit for small $P^*$. 

The effective pressure approach to the yield criterion of wet grains ignores such sophistications, as well as density or microstructural changes due to capillary forces.
It assumes critical states to be in correspondence for different values of $P^*$, as though the introduction of capillary forces, pushing grains 
against their neighbors,  were equivalent to the application of a larger confining pressure. Such a crude approach is in fact surprising successful as a rough approximation, to
predict the increase of $\mu^*$ for decreasing, but not too small, $P^*$, say $P^*\ge 1$ (below $P^*=1$, the increase of $\mu^*$ is overestimated). The effective pressure might
be evaluated upon adding, to the applied pressure, the capillary contribution to the average stress. This contribution might itself be estimated from density and coordination numbers, 
which leads to a Mohr-Coulomb form \eqref{eq:mohrcoulomb} for the variation of tensile strength $\mu^*_0 \sigma_{22}$ with normal stress  $\sigma_{22}$ (if $\Phi$ and $z$  do not vary too much).  Such a form of the critical state plasticity criterion 
proves however inadequate to describe the whole range of reduced pressures $P^*$: data are incompatible with a $P^*$--independent macroscopic cohesion $c$. 

Another remarkable feature of the measured rheology, compared to dry granular materials, is the slow variation with $I$  of both macroscopic rheological parameters 
such as $\mu^*$, and microscopic data such as coordination numbers. Although we could not observe a direct correlation, a slower increase of function $\mu^*(I)$ could signal
a shear banding tendency. 

Many of those rheophysical features are explained by, or, at least, related to, the strong clustering tendency emerging as attractive forces gradually become the dominant ones, upon decreasing $P^*$.  As contacts are stabilized by attractive forces, they do not  so easily open as the network is being sheared. When they do (which happens preferentially in 
the extension direction within the average shear flow, whence a fabric anisotropy of distant interactions contributing to shear strength), the network of grains bonded by
liquid bridges might still be connected, forming enduring connected clusters. The survival of such clusters over quite notable strain intervals (reaching several unities with sizable probability) should limit the dilating tendency of faster flows. It also maintains a network in which the capillary forces act in closer similarity to an effective pressure. 

Admittedly, this is still a descriptive rheophysical scenario. More quantitative studies should be carried out to better characterize the deformation mechanisms of the grain clusters.
The shear banding phenomenon certainly deserves detailed investigations, in which sample size and shape effects should be systematically assessed, and partly localized 
velocity an density fields analyzed and related to a stability analysis. 

Although we argued that our measurements agree with the limited available experimental results, more laboratory data should certainly be used, with, if possible both information on
rheology and on micromorphology and liquid distribution. 

Finally, a complete numerical model should eventually be designed, capable of dealing with saturations exceeding the limited pendular range, and of describing the liquid motion. 
One should thus be equipped to model the mixing process as well as the rheology of the homogeneous mixture of the grains and the liquid.
The Lattice Boltzmann method for a diphasic interstitial fluid medium, coupled to a DEM description of grain motion, is a promising perspective~\cite{DeRiRa15,Roux2015}.


\end{document}